\documentclass[12pt,preprint]{aastex}

\usepackage{graphicx}
\usepackage{txfonts}
\usepackage{natbib}
\shorttitle{Time variations of the observed H$\alpha$ line profiles....}
\shortauthors{Falewicz et al.}

\begin{document}
%


\title{Time variations of observed H$\alpha$ line profiles and precipitation depths of non-thermal electrons in a solar flare}


\author{Robert Falewicz\altaffilmark{1}}
\affil{$^{1}$ Astronomical Institute, University of Wroc{\l}aw,
51-622 Wroc{\l}aw, ul. Kopernika 11, Poland}
\email{falewicz@astro.uni.wroc.pl}

\author{Krzysztof Radziszewski\altaffilmark{1}}
\email{radziszewski@astro.uni.wroc.pl}
\author{Pawe{\l} Rudawy\altaffilmark{1}}
\email{rudawy@astro.uni.wroc.pl}
\and

\author{Arkadiusz Berlicki\altaffilmark{1,2}}
\affil{$^{2}$ Astronomical Institute, Academy of Sciences of the
Czech Republic, Ondrejov}
\email{berlicki@astro.uni.wroc.pl}




\begin{abstract}

We compare time variations of the H$\alpha$ and X-ray emissions observed during the pre-impulsive and impulsive phases of the C1.1-class solar flare on 21 June 2013 with those of plasma parameters and synthesized X-ray emission from a one-dimensional hydro-dynamic numerical model of the flare. The numerical model was calculated assuming that the  external energy is delivered to the flaring loop by non-thermal electrons. The H$\alpha$ spectra and images were obtained using the Multi-channel Subtractive Double Pass spectrograph with a time resolution of 50~ms. The X-ray fluxes and spectra were recorded by the {\it Reuven Ramaty High-Energy Solar Spectroscopic Imager} ({\it RHESSI}). Pre-flare geometric and thermodynamic parameters of the model and the delivered energy were estimated using {\it RHESSI} data.

The time variations of the X-ray light curves in various energy bands and the those of the H$\alpha$ intensities and line profiles were well correlated. The time scales of the observed variations agree with the calculated variations of the plasma parameters in the flaring loop footpoints, reflecting the time variations of the vertical extent of the energy deposition layer. Our result shows that the fast time variations of the H$\alpha$ emission of the flaring kernels can be explained by momentary changes of the deposited energy flux and the variations of the penetration depths of the non-thermal electrons.

\end{abstract}




\section{Introduction} \label{sec:intro}

Observations of solar flares in hard X-rays (HXR) reveal bright, compact emission sources at flaring loop footpoints, where dense and cold chromospheric plasma is heated by beams of non-thermal electrons (NTEs). In the light of strong chromospheric lines such as the hydrogen H$\alpha$ line ($\lambda=6562.8$~\AA), the flare emission sources are in the form of bright localized kernels as well as ribbons. Similarities of the light curves of the HXR and H$\alpha$ sources are strikingly similar \citep{Radziszewski2007,Radziszewski2011,Radziszewski2013}. The NTEs are accelerated in so-called primary energy release sources, located close to the tops of flaring loops. During the precipitation along the magnetic loops, the NTEs are slowed down and thermalized by Coulomb collisions with ambient ions, mostly in the relatively dense plasma near the loop footpoints, but a small part of their kinetic energy is radiated as HXR bremsstrahlung. The energy delivered by the NTEs also powers heating and macroscopic motions of the upward-moving, ``evaporated'' plasma, and emission in soft X-rays and chromospheric lines \citep{Brown1971,Antonucci1984,Antonucci1999,Fisher1985a,Fletcher2011,Holman2011}. The parameters of the NTE beams, like the distribution and total energy, and precipitation depths vary rapidly, with time-scales of seconds. The time variations of the HXR emission are in addition a function of the temporal and spatial variations of the plasma properties along the flaring loop (mostly the density and the temperature of the plasma near the loop footpoints: see \cite{Battaglia2012}).

Strong chromospheric emission lines in flares, particularly the H$\alpha$ line, are formed at heights extending from the upper photosphere to the upper chromosphere \citep{Vernazza1971,Vernazza1981,Kasparova2002,Berlicki2004}. The intensities and line profiles of the H$\alpha$ emission depend on many factors, e.g. the vertical and horizontal stratification of plasma characteristics inside the loop \citep{Fisher1985a,Heinzel1994,Allred2005,Berlicki2007,Falewicz2015}, the bulk and turbulent plasma motions \citep{Abbett1999,Berlicki2005}, the precipitation time of the NTEs \citep{Kasparova2002}, the propagation velocities of the conduction fronts \citep{Reep2016}, the properties and variations of the NTE beams penetrating the plasma at the loop footpoints \citep{Kasparova2009}.

Physical processes causing the abrupt heating of plasma during solar flares and the subsequent evolution of the hot plasma within flaring loops are not fully understood. A comparison of the observed features of the flaring plasma with results from numerical models should help in a better understanding of the physics of the flaring plasma. Considering all the factors discussed above, the analysis of the time variations of the observed X-ray and H$\alpha$ emissions could deliver irreplaceable data about the time variations of the spatial distributions of plasma parameters and processes in the flaring loops. However, due to very short time scales of the variations of the NTEs beams, the X-ray bremsstrahlung emission, and the H$\alpha$ line emission, the applied observational data should be collected with a sufficiently high time resolution.

In this paper we compare time variations of the H$\alpha$ and X-ray emissions observed during the pre-impulsive and impulsive phases of a C1.1-class solar flare in the NOAA~11772 active region on 21 June 2013 with the variations of plasma parameters in the one-dimensional hydro-dynamic (1D-HD) numerical model of the flare and with the variations of synthesized X-ray emission. The 1D-HD model was calculated using a modified hydrodynamic one-dimensional Solar Flux Tube Model developed at the Naval Research Laboratory \citep[see][for details]{Mariska1982,Mariska1985,Falewicz2009a}, under the assumption that energy is delivered to the flaring loop by the NTEs. Pre-flare parameters of the model and energy fluxes delivered to the plasma during the flare were estimated using data from the {\it Reuven Ramaty High Energy Solar Spectroscopic Imager} ({\it RHESSI}: \cite{Lin2002}). The H$\alpha$ line profiles and light curves, measured at several wavelengths over the H$\alpha$ line profile, were recorded with high time resolution (0.05~s) using the Multi-Channel Subtractive Double Pass spectrograph (MSDP) which was installed at the focus of the Horizontal Telescope at the Bia{\l}k{\'o}w Observatory of the University of Wroc{\l}aw, Poland \citep{Mein1991,Rompolt1994}.

In the following, Section 2 describes the flare observations and data reduction, Section 3 the numerical model of the flare, and Section 4 the flare's H$\alpha$ emission. The conclusions and discussion are given in Section 5.

\section{C1.1 Solar Flare on 21 June 2013} \label{sec:flare}

\begin{figure}[t]
\figurenum{1}
\begin{center}
\includegraphics[width=11.0cm]{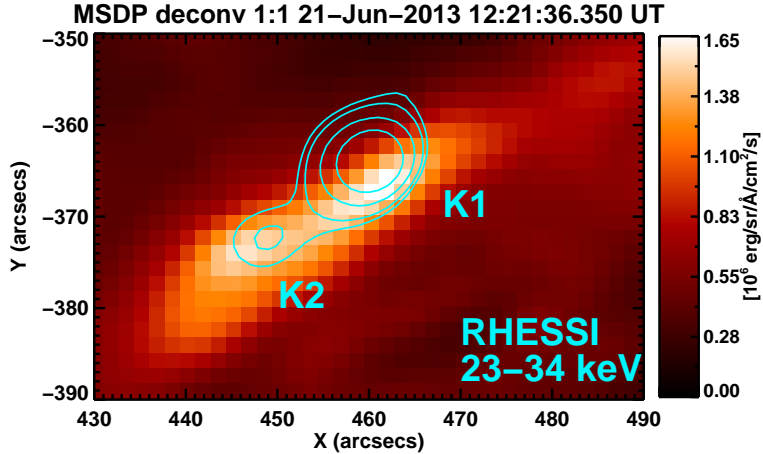}
\end{center}
\vspace{-1.0 cm}
\caption{\small The H$\alpha$ emission (image) and  the {\it RHESSI} HXR 23--34~keV emission (blue contours at 16.5{\%}, 20{\%}, 30{\%} and 50{\%} of the maximum signal) of the C1.1 \textit{GOES}-class solar flare on 21 June 2013 (12:21:36~UT). The {\it RHESSI} integration time was 64~s. Two bright H$\alpha$ flaring kernels are marked K1 and K2. North is up.
\label{Fig01}}
\end{figure}

\begin{figure}[t]
\figurenum{2}
\begin{center}
\includegraphics[width=9.0cm]{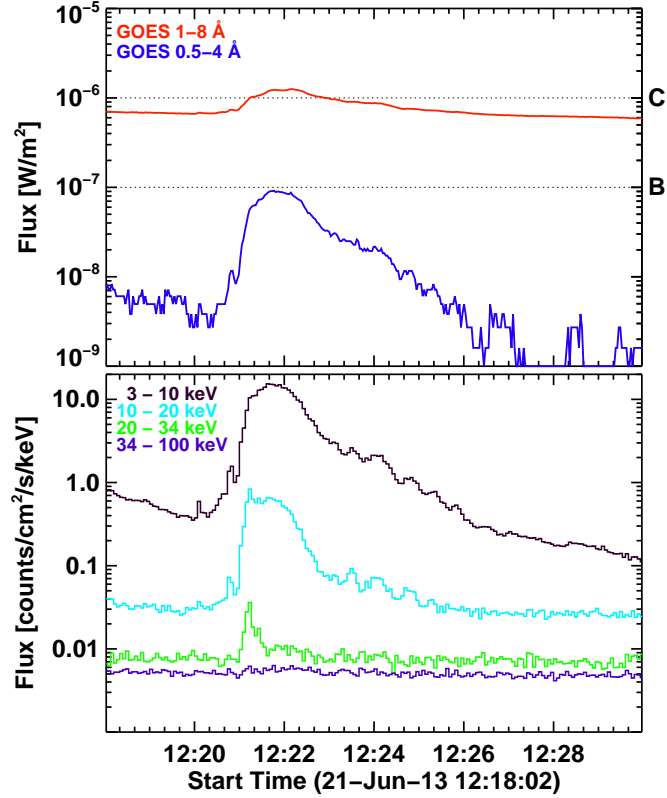}
\end{center}
\vspace{-1.0 cm}
\caption{\small \textit{GOES} X-ray 0.5--4~{\AA} and 1--8~{\AA} light curves (upper panel) and the {\it RHESSI} light curves in four energy bands: 3--10~keV, 10--20~keV, 20--34~keV, and 34--100~keV (lower panel) during the 21 June 2013 flare between 12:18~UT and 12:30~UT.
\label{Fig02}}
\end{figure}

The C1.1 \textit{GOES}-class solar flare occurred in a complex, $\beta\delta$ magnetic-type active region, NOAA~11772  (coordinates x=~457'', y=~-368''; S15W31) on 21 June 2013 at 12:20~UT (Fig.~\ref{Fig01}). The 1--8~{\AA} X-ray emission of the flare recorded by {\it GOES-15} reached maximum at 12:22~UT, returning to the pre-flare background level at about 12:27~UT (Fig.~\ref{Fig02}, upper panel). The impulsive phase of the flare started at 12:21:00~UT, reaching maximum 12--14 seconds later (Fig.~\ref{Fig02}, lower panel). The increase in the soft X-ray (SXR) emission ($< 12$~keV) recorded by {\it RHESSI} was simultaneous with that in {\it GOES-15} 0.5--4~{\AA} emission.

Significant HXR emission was recorded by {\it RHESSI} up to 34~keV only. The X-ray fluxes emitted by the flare were so low that the {\it RHESSI} attenuators were not activated during the flare (i.e., they remained in the "A0" state). The signals were summed over the front segments of seven of the nine detectors: 1F, 3F, 4F, 5F, 6F, 8F, and 9F. The recorded data were analysed using the OSPEX software of the SolarSoftWare (SSW) package. X-ray spectra were obtained with time resolution of four seconds and binned in the energy bands dE = 0.3~keV in the 4--15 keV energy range, and dE = 1.0 keV in the 15--100 keV energy range. The pulse pileup, decimation, and albedo effects (due to Compton back-scattering from the photosphere \citep{Bai1978, Kontar2010}) were corrected using the \emph{RHESSI} standard analysis tool. The full two-dimensional detector response matrix was applied to convert count rates to photon fluxes. The accumulation time was increased to 20~s before the flare impulsive stage in order to maintain positive count rates (particularly in the 6--25 keV energy range).

The fluxes recorded by \emph{RHESSI} in the 6--10~keV band during the flare were slightly enhanced by a background that slowly decreased during the flare, while the fluxes in the 10-20~keV and 20-34~keV energy bands were enhanced by steady background fluxes. Above 34~keV the HXR background dominates over the emission of the flare, so the fluxes above this energy were not analyzed. In the case of the \emph{RHESSI} data, a background is usually modeled as a linear or higher-order polynomial interpolation between the averaged fluxes recorded before and after the flare. As the time period investigated here is very short, a linear fit to the background rates before and after the flare is adequate for our purposes. For energies $<12$~keV, the linear fit was to background rates at 12:11~UT and 12:30~UT; for energies $>12$~keV, the corresponding times were 12:20~UT and 12:29~UT.  The background fluxes were also removed from \emph{GOES} data using a linear fit of fluxes before and after the flare.

\begin{figure}[t]
\figurenum{3}
\begin{center}
\includegraphics[width=5.0cm]{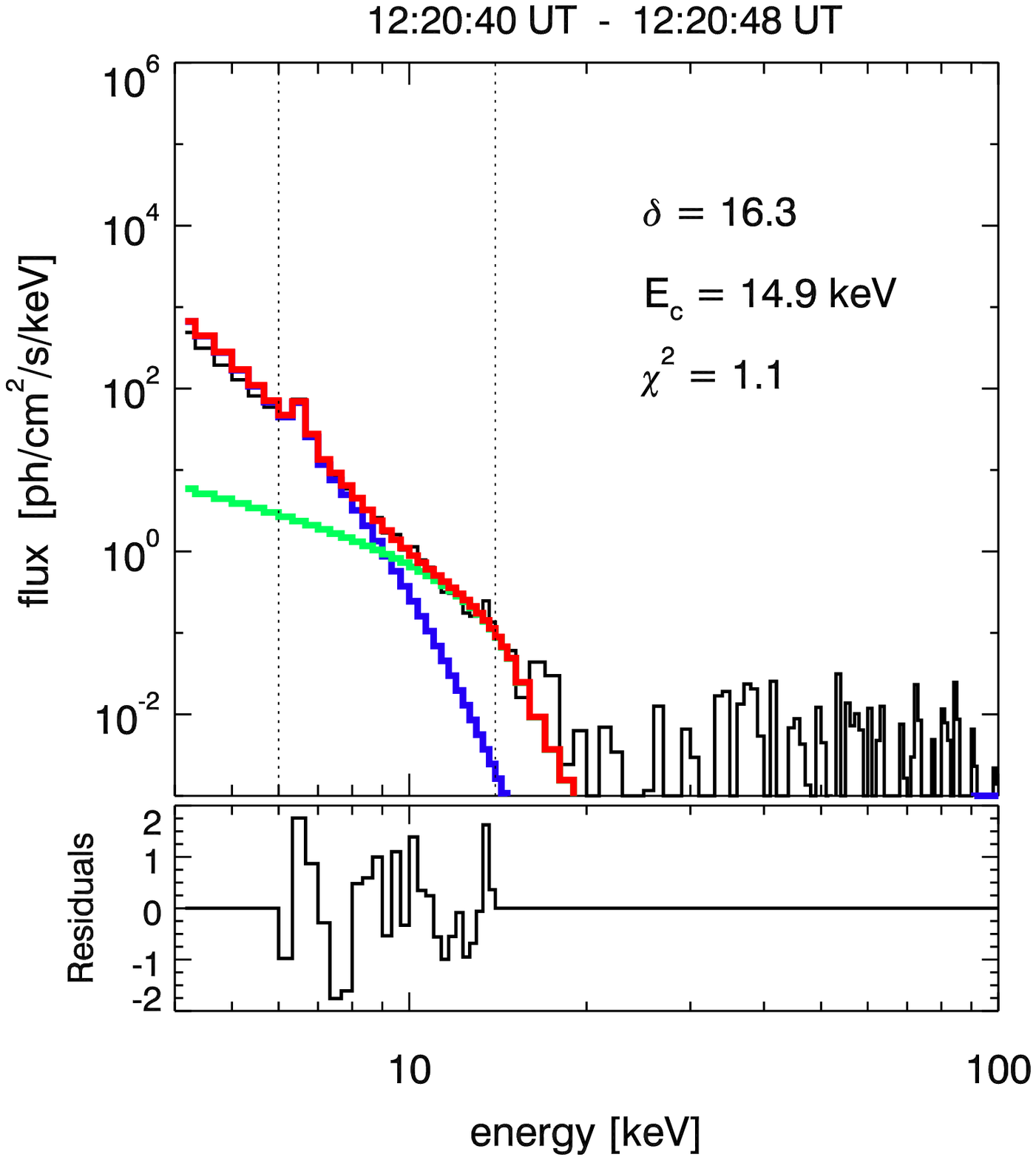}
\includegraphics[width=5.0cm]{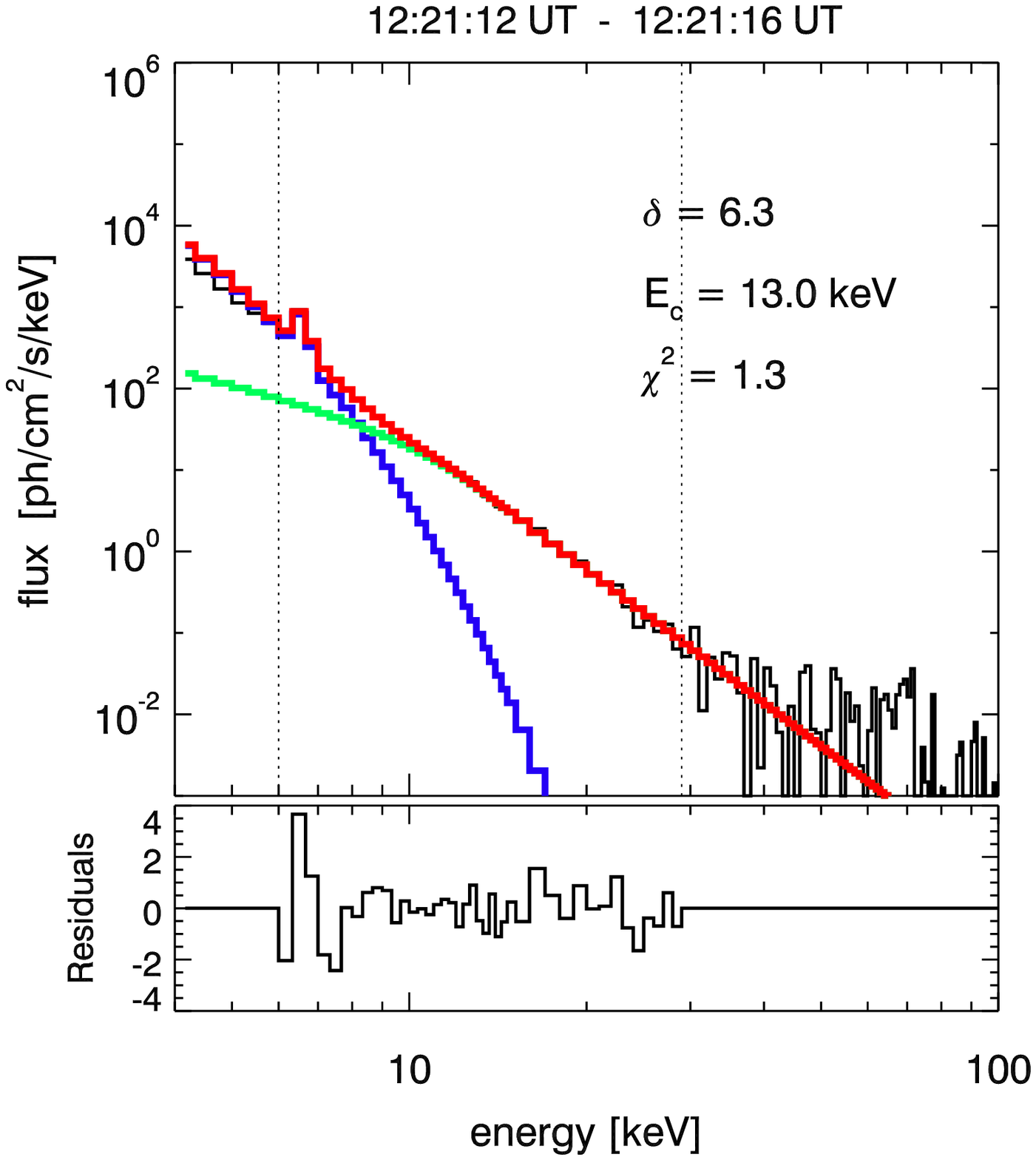}
\includegraphics[width=5.0cm]{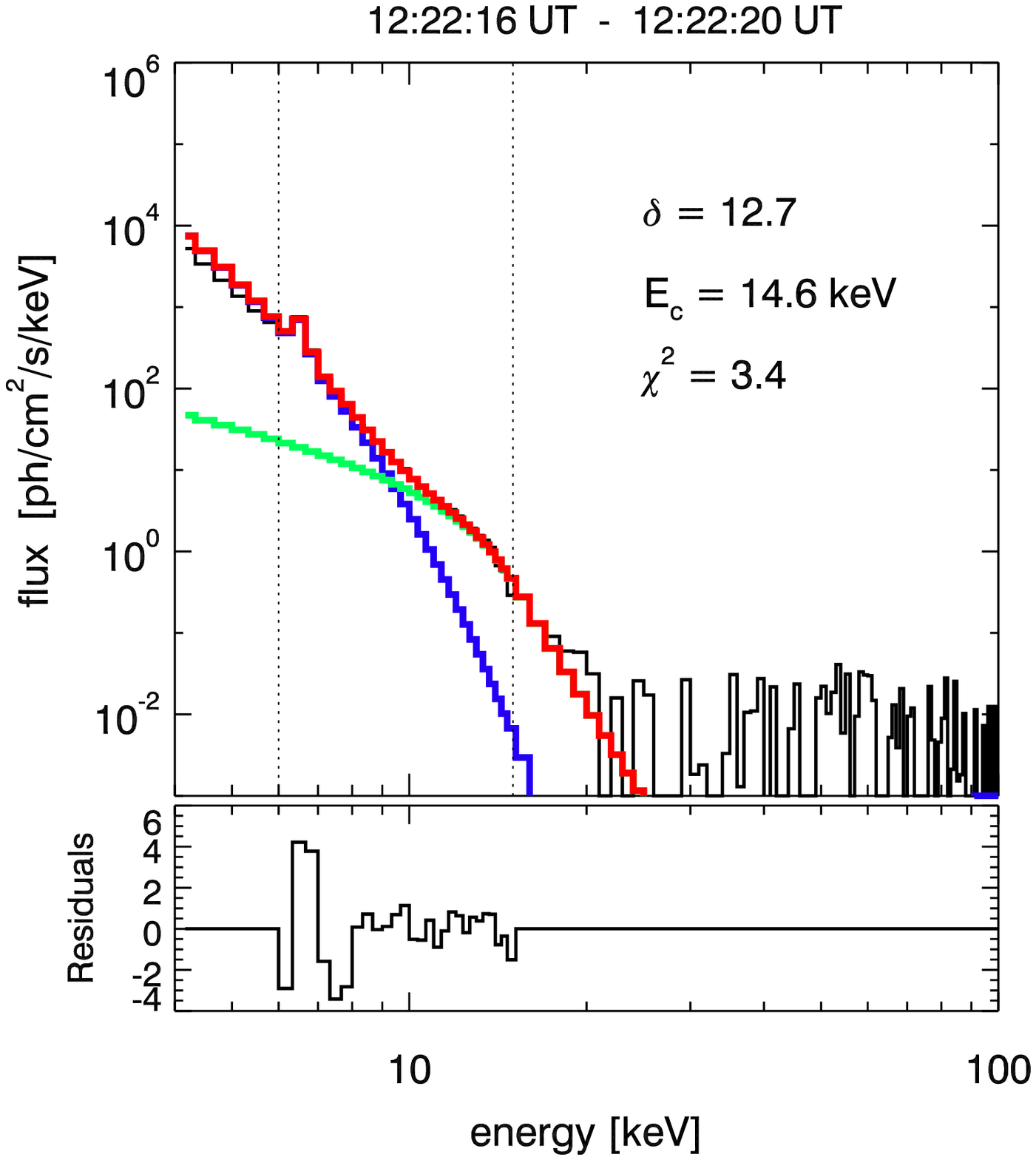}
\end{center}
\vspace{-0.1 cm}
\caption{\small \emph{RHESSI} spectra recorded during the pre-impulsive phase (left panel), the impulsive phase (middle panel), and at the end of the impulsive phase (right panel) of the 2013 June 21 flare. The spectra were fit with the single-temperature thermal model (blue line) and thick-target model (green line). The total fitted spectra are shown as red lines. The obtained values of $\delta$ and $\rm E_c$ were determined based on the thick-target model and used as the parameters of the injected NTEs beams in the numerical model of the flare. Energy ranges used for fitting each spectrum are shown with vertical lines. Residuals for the fits are shown below the spectra.
\label{Fig03}}
\end{figure}

During the impulsive phase, {\it RHESSI} spectra show both thermal and (at higher energies) non-thermal components. \emph{RHESSI} spectra were fit using a forward and backward automatic fitting procedure, starting from a maximum of the impulsive phase, when the nonthermal component was strong and clearly visible. Automatically evaluated parameters of the fit of the thermal and non-thermal components were controlled and corrected when the time variations were too rapid. The correction procedure consisted of four steps. First, the corrected parameter was fixed using a value selected to obtain a first-guess fit of the observed spectrum. Second, all remaining parameters of the fit were optimized automatically. Third, the corrected parameter was freed and the automatic fit of the spectrum was repeated. At the end of the correction procedure,  values of $\chi^{2}$ and residuals were controlled. Generally, all the fitted parameters evolved in a quasi-continuous manner and the described procedure of the parameter correction was applied to the spectra recorded during the beginning of the flare only. We applied the ``single temperature thermal plus thick-target version 2'' model fits (vth + thick2) in the {\it RHESSI} analysis software. The thick-target model (version~2) was defined by the total integrated electron flux $N_{nth}$, the power-law index of the electron energy distribution $\delta$, and the low-energy cutoff of the electron distribution $E_c$. Figure~\ref{Fig03} shows three examples of the spectra taken during the pre-impulsive and impulsive phases of the flare as well as just after the impulsive phase.

\begin{figure}[t]
\figurenum{4}
\begin{center}
\includegraphics[width=8 cm]{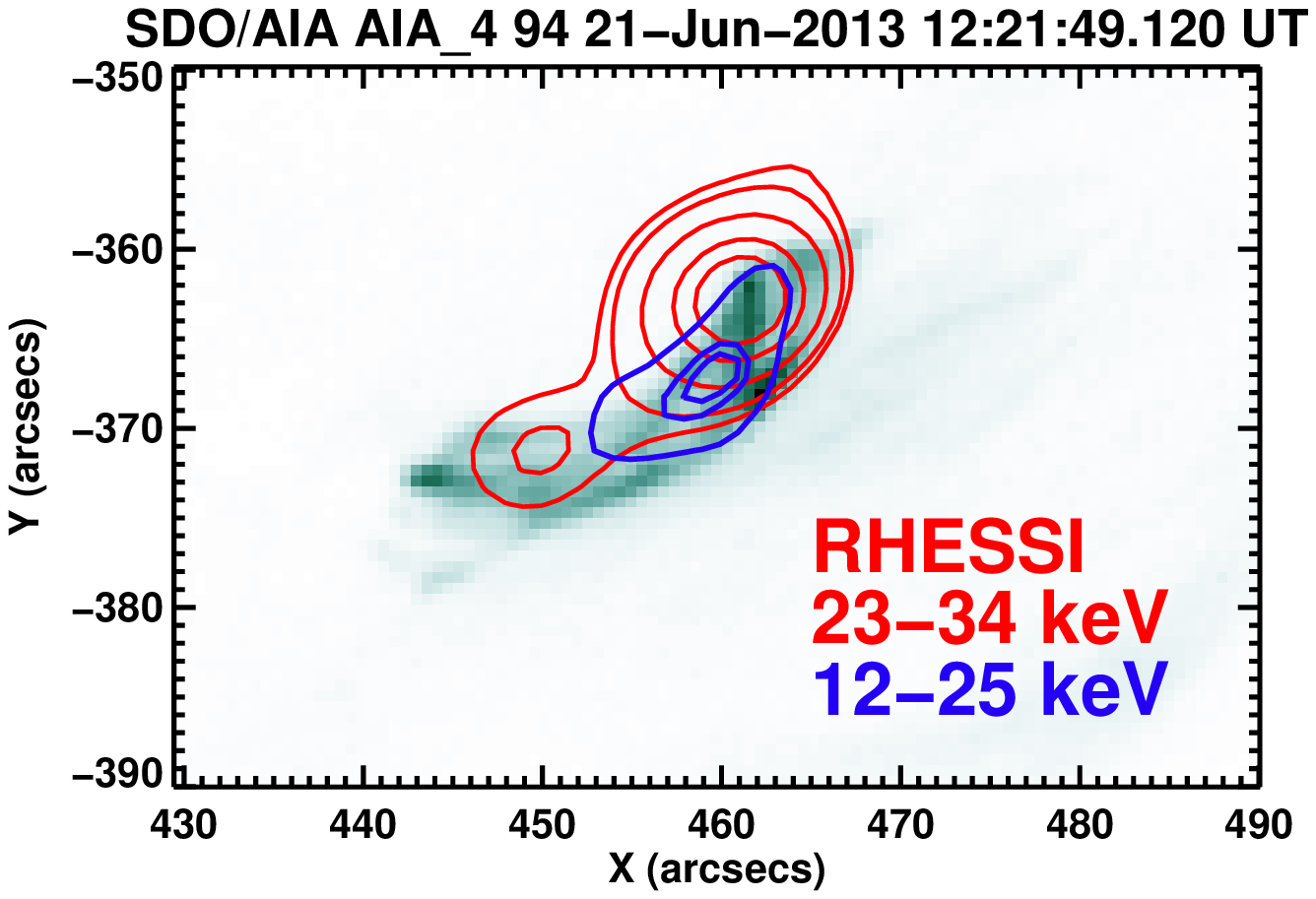}
\includegraphics[width=8 cm]{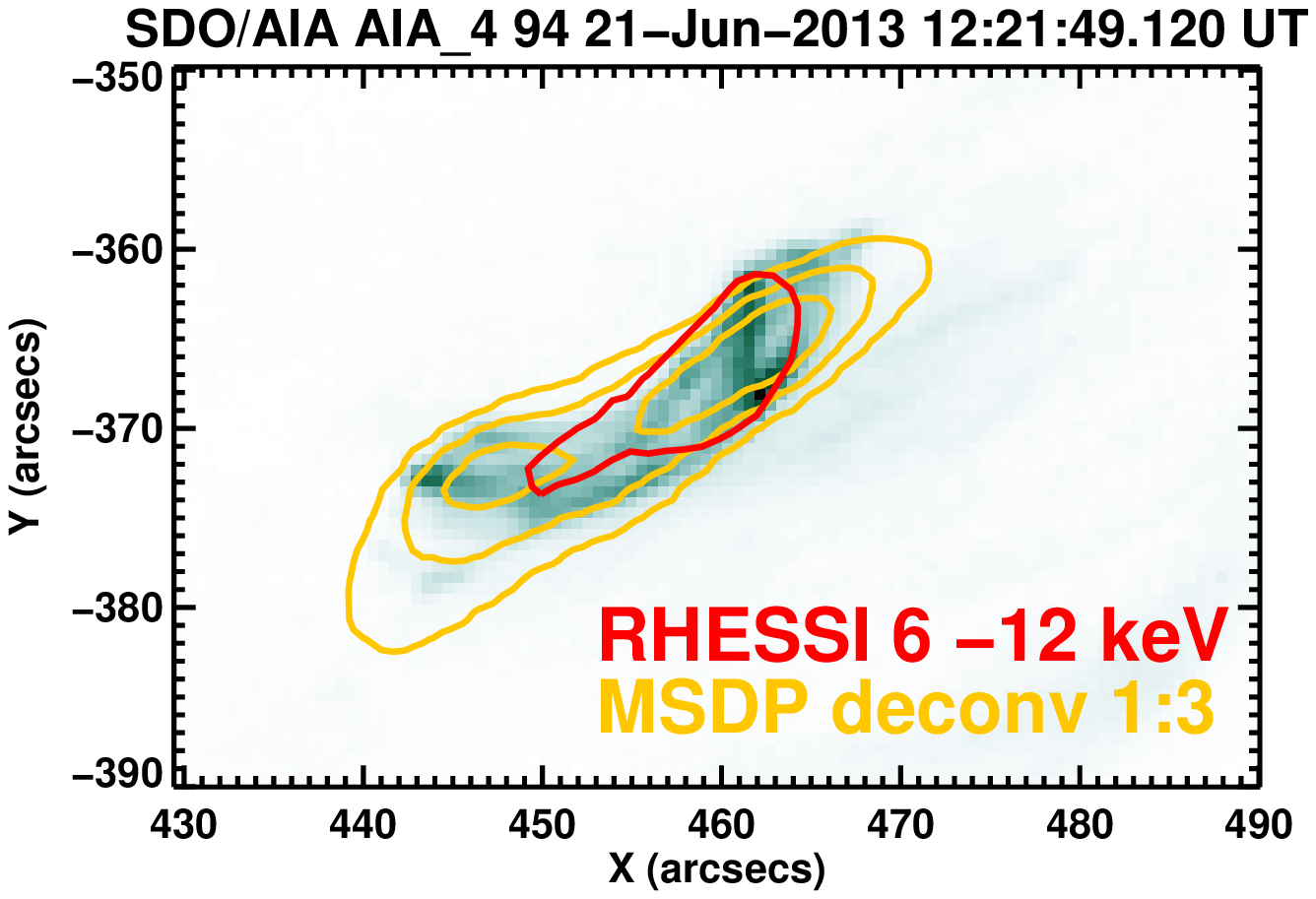}
\end{center}
\vspace{-1.0 cm}
\caption{\small Left panel: co-aligned {\it SDO}/AIA 94~{\AA} (image), and {\it RHESSI} hard X-ray 12--25~keV and 23--34~keV (blue and red contours, respectively) emissions of the 21 June 2013 flare. The {\it RHESSI} 23--34~keV contours are displayed at 16.5{\%}, 20{\%}, 30{\%}, 50{\%} and 70{\%} of maximum signal, the {\it RHESSI} 12--25~keV contour is drawn at 50{\%}, 80{\%}, and 90{\%} of maximum signal. Right panel: co-aligned {\it SDO}/AIA 94~{\AA} (image), MSDP/H$\alpha$ (yellow contours), and {\it RHESSI} 6-12~keV (red contour) emission. The H$\alpha$ contours are drawn at 70{\%}, 80{\%}, and 90{\%} of maximum of the signal. The {\it RHESSI} 6--12~keV contour is drawn at 50{\%} of maximum signal. The integration time of the {\it RHESSI} fluxes was 64~s, and the fluxes were recorded between 12:21:06~UT and 12:22:10~UT. The MSDP image was taken at 12:21:36.35~UT.
\label{Fig04}}
\end{figure}

X-ray images of the flare were constructed using {\it RHESSI} data collected with the sub-collimators 3F, 4F, 5F, 6F, 8F, and 9F. The PIXON imaging algorithm was applied with the effective pixel size of one arc second. Photons were integrated in time steps of 8~s to obtain adequate signal-to-noise ratios \citep{Metcalf1996,Hurford2002}. The energy ranges of 6--12~keV, 12-25~keV, and 23-34~keV were selected in order to best portray the flare structures. The images restored in the 6--12~keV energy range (see Fig.~\ref{Fig04}, right panel) show an elongated bright structure, located between two H$\alpha$ brightenings seen in the chromosphere. The 12--25~keV images (see Fig.~\ref{Fig04}, left panel) show a compact emission source, which coincided with the central part of an extended structure visible in the 23--34~keV energy band. The magnetograms obtained with the {\it SDO}/HMI instrument \citep{Scherrer2012} were used to determine magnetic polarities in the flare region, providing an additional confirmation of the proper selection of the loop footpoints. The opposite ends of the extended structure in the 23--34~keV energy band were located in the regions having opposite magnetic polarities. Images recorded at the same time by the {\it SDO}/AIA instrument \citep[see][]{Lemen2012} in the 94~{\AA} band show an extended and multi-threaded structure, similar to the elongated emission source recorded in H$\alpha$. Images recorded by {\it SDO}/AIA in the 171~{\AA} band between 12:19:50~UT and 12:23:01~UT clearly show that the flare occurred in a part of a complex system. Owing to the broad wavebands of the {\it SDO}/AIA EUV filters, a full interpretation of the revealed structures is challenging.

The main geometric characteristics of the flaring loops were evaluated using the {\it RHESSI} images. The effect of the perspective foreshortening was reduced using the method of \citet{Aschwanden1999}. The cross-section of the loop was assumed to be constant along the loop and it was estimated as a mean area of both loop footpoints, arbitrarily delimited using 30$\%$ contours of their local maximum fluxes in the 20--34~keV range. The areas and positions of the centroids of the footpoints were calculated using the SSW \emph{CENTROID} procedure. The resulting cross-section of the loop was equal to S~=~(1.65~$\pm$~1.04)~$\times$~10$^{17}$~cm$^{2}$. The 30$\%$ contour was selected as the delimiter of the footpoints in accordance with our previous work \citep[see][]{Falewicz2011, Falewicz2014, Falewicz2015} in order to moderate the mean density of the energy flux in the loop, while in the 1D-HD numerical models the energy is formally delivered uniformly over the whole cross-section of the loop. The half-length of the loop equal to L$_0$~=~(8.81~$\pm$~1.13)~$\times$~10$^{8}$~cm was estimated as a distance between centroids of the feet, under assumption that the loop was semi-circular.

Assuming a perpendicular position of the semi-circular flaring loop on the solar surface, the apparent positions of the centroids of the footpoints observed in various energy bands can be easily converted into heights above the solar surface. In the case of the brighter, north-west footpoint of the flaring loop during the impulsive phase, the centroids were shifted by $\Delta$s~=~1.7~{\it arcsec} when observed in the slightly overlapped 15--26~keV and 23--34~keV energy bands. The formal errors of the calculated positions of the centroids are about $\delta$=1.5~{\it arcsec}. The measured angular distance of the centroids is equivalent to a vertical extent of $\Delta$h~=~2550~km. The relatively low energy fluxes emitted by the flare prevented an application of images restored in the non-overlapping energy bands and the evaluation of the vertical extents before and after the maximum.

The flare was also observed in the Bia{\l}k{\'o}w Observatory of the University of Wroc{\l}aw, Poland, using the MSDP spectrograph and the Horizontal Telescope (HT). The HT has a Jensch-type coelostat with two 30-cm flat mirrors and a 15-cm f/27 achromatic main objective. The entrance window of the MSDP spectrograph covers an area of $942\times119~{\it arcsec}^{2}$ on the Sun when illuminated by the HT. The 2D spectra--images are formed with a nine-channel prism box and are recorded with an Andor iXon3 885 camera ($1002\times1004~px^{2}$, $1~px^{2}= 1.6~{\it arcsec}^{2}$). The time cadence was 20~images per second during the observations. Raw observations were reduced using standard reduction procedures. Extended information concerning the high time resolution spectra--imaging observations and data reduction were given previously  \citep{Radziszewski2006,Radziszewski2007,Radziszewski2011,Radziszewski2013}. For each spectra--image we calculated: (i) H$\alpha$ line spectra in the range of $\Delta\lambda=\pm 1.2$~{\AA} from the H$\alpha$ line center ($\lambda=6562.8$~\AA, marked next as H$\alpha_0$) for all pixels of the FOV, (ii) the quasi-monochromatic images of the entire region, reconstructed in 13 wavebands 0.06~\AA~wide each, and separated by 0.2~{\AA} from each other over the H$\alpha$ line profile, and (iii) the light curves of selected flaring kernels in various wavelengths over the range $\Delta\lambda=\pm 1.2$~{\AA} from the central wavelength (H$\alpha_0$).

Ground-based observations of the Sun are always affected by a variable atmospheric seeing, causing deformations of apparent shapes and variations of the brightness of recorded structures. The collected data are also influenced by telescope guiding errors. Special procedures were applied to correct all these effects \citep[see][for details]{Radziszewski2007}. The Richardson-Lucy deconvolution algorithm was also applied to enhance the quality of the resultant MSDP images \citep{Richardson1972,Lucy1974}. The continuum filtergrams taken with the {\it SDO}/HMI telescope were used for the exact alignment of the H$\alpha$ ground-based and satellite observations.

The time resolution of the X-ray light curves recorded with {\it RHESSI} is limited to four seconds by the spacecraft spin. For a qualitative comparison of the observed X-ray and H$\alpha$ light curves, X-ray data with 0.25~s time resolution were obtained using the demodulation program prepared by Hurford and available in the SSW package \citep{Hurford2004}. To be consistent with our previous papers, the observed time variations of the H$\alpha$ light curves were compared with the time variations of the X-ray light-curves measured for slightly altered energy bands: 3--10~keV, 10--20~keV and 20--34~keV \citep[see \emph{e.g.}][]{Radziszewski2007,Radziszewski2011}.

\section{Numerical Model of the Flare} \label{sec:model}

A solar flaring loop is a three-dimensional object possibly having a complex, multi-threaded internal structure. Ground-based and satellite-based instruments are not able to resolve the internal structure of flaring loops as well as heating episodes and thermodynamic evolution of the individual threads. The flaring loops are also surrounded by a dynamic environment of a solar active region, dominated by evolving and restructuring magnetic fields. Many basic parameters of the numerical models have to be approximated or freely chosen due to the incompleteness of the available observational data. For example a number of the internal threads (i.e. the filling-factor of the observed macroscopic loop) and individual lengths of the threads has to be chosen, while the short flux tubes evolve much faster than the long ones due to an influence of the thermal conductivity and the mean plasma parameters of a set of small flux tubes \citep[see \emph{e.g.}][and papers there referred]{Klimchuk2008}. The results of the calculations depend also on an adapted model of an energy deposition mechanism inside the threads. For these reasons, substantial differences between evolutions of the observed and numerically modeled flaring loops are unavoidable despite the complexity of the codes presently applied. To solve this complex problem, \citet{Warren2006} elaborated a multi-threaded and time-dependent model of the so-called Masuda flare observed on January 13, 1992. He showed that the discrepancies between the simulations and observations may be reduced using a multi-threaded 1D-HD model, but only by approximating the numerous auxiliary parameters, e.g. the number of threads or the time-sequence of the energy deposition episodes inside individual threads. \citet{Falewicz2015} investigated a 2D model of the flaring loop having a continuous distribution of the parameters of the plasma across the loop and powered by a variable energy flux. This model was also compared with the relevant 1D-HD models. They found that multi-threaded and multi-dimensional numerical models of the flare with correctly selected spatial distributions of the input energy and a realistic physical model of the loop can produce results similar to the results of the 2D models, as well as that 1D-HD models ensure general correctness of the obtained results and approximate real flaring structures. The 2D or even 3D time-dependent numerical models of the flaring loops basically allow many important properties of the flares to be accounted for, like the internal structure of the magnetic field inside the observed flaring loops, interaction between the threads, and the individual time-sequences of the energy deposition episodes in the individual threads. However, the observational limitations imposed by the presently available solar instruments do not allow any qualitative measurements of these characteristics, and the computation times of the 2D (or 3D) models are  significantly greater than those of the 1D models. Thus, taking into account all limitations of the 1D-HD models, they still remain a valuable tool in investigations of the solar flares thanks to their moderate complexity and moderate volume of the necessary calculations. The 1D-HD models have also bee applied with success in the analysis of flares leading to some understanding of the physics of the plasma inside flaring loops \citep[see \emph{e.g.}][and many other]{Mariska1985, Fisher1985a,Fisher1985b, Fisher1985c, Mariska1989, Serio1991, Reale1997, Siarkowski2009, Falewicz2009a, Falewicz2009b, Falewicz2011, Falewicz2014}.

The 1D-HD numerical model of the flare on June 21, 2013 was calculated using the modified hydrodynamic one-dimensional Solar Flux Tube Model \citep[see][for details]{Mariska1982,Mariska1989,Falewicz2009a}. It was assumed that the flare plasma was heated only by time-variable NTE beams, as has been found for some flares by Falewicz et al. (2011, 2014). However, this may not always be the case, and some additional energy source is needed \citep{Liu2013,Aschwanden2016}.

The numerical model of the flaring loop was prepared using numerous specific geometric and thermodynamic parameters. The loop length and cross section, the initial pressure at the transition region and plasma temperature were estimated from {\it RHESSI} observations. To obtain agreement between synthetic and observed X-ray light curves, the radius of the loop was refined arbitrary in a range of one {\it RHESSI} pixel (it is $\pm$~725~km) and the cross section of the loop applied in the numerical model was equal to S~=~1.01~$\times$~10$^{17}$~cm$^{2}$. The applied cross-section agrees well with an area of the structures observed by the SDO/IAI telescope in 94~{\AA} band. A constant magnetic field strength of 200~G was assumed, following estimates of \citet{Schmeltz1994} and \citet{Aschwanden2005}.

The observed and synthesized X-ray light curves calculated for the 1D-HD model of the flare were compared in three energy bands: 6--10~keV, 10--20~keV and 20--34~keV. The synthesized X-ray fluxes were calculated in photons/cm$^{2}$/s/keV. The observed X-ray fluxes were converted to the photons/cm$^{2}$/s/keV unit using the instrumental response matrix with diagonal coefficients only. Thus, the secondary factors influencing {\it RHESSI} data and described by the off-diagonal elements of the response matrix, like the K-escape and the Compton scattering, were not taken into account. The applied method is valid because all investigated energy bands lay above the count-rate peak of the analyzed flare, which was equal to about 6 keV (the attenuators were not activated during the whole event and remained in A0 state) while off-diagonal contributions are significant only below the count-rate peak. The method used introduces an insignificant systematic error of a few percent only. Time variations of the {\it GOES} 1--8~{\AA} and {\it GOES} 0.5--4~{\AA} light curves were also computed. The detailed description of the method used for calculation of the \emph{GOES} light curves can be found in \citet{Falewicz2014}.

The steady-state spatial and spectral distributions of the NTEs in the loop were calculated for each time step of the model using a Fokker-Planck model \citep{McTiernan1990}. The input data consisted of an injected energy flux, an isotropic distribution of NTEs at the top of the loop, a model of a distribution of the plasma parameters along the flaring loop, and a model of the magnetic field. Spatial distributions of the thermodynamic parameters of plasma, X-ray thermal and non-thermal emissions, and the integral fluxes in the selected energy ranges were calculated. The thermal emission measure of the optically thin plasma was based on the X-ray continuum and line emission calculated using the CHIANTI (version 7.1) atomic code \citep{Dere1997,Landi2006}. For the plasma temperatures above 10$^{5}$ K the element abundances are based on a coronal abundance set \citep{Feldman2000}, while below 10$^{5}$ K photospheric abundances were applied. To compare the synthesised and the observed fluxes, the calculated thermal and non-thermal emission and the estimated observed background emission were summarized.

\begin{figure}[t]
\figurenum{5}
\begin{center}
\includegraphics[width=8.0cm]{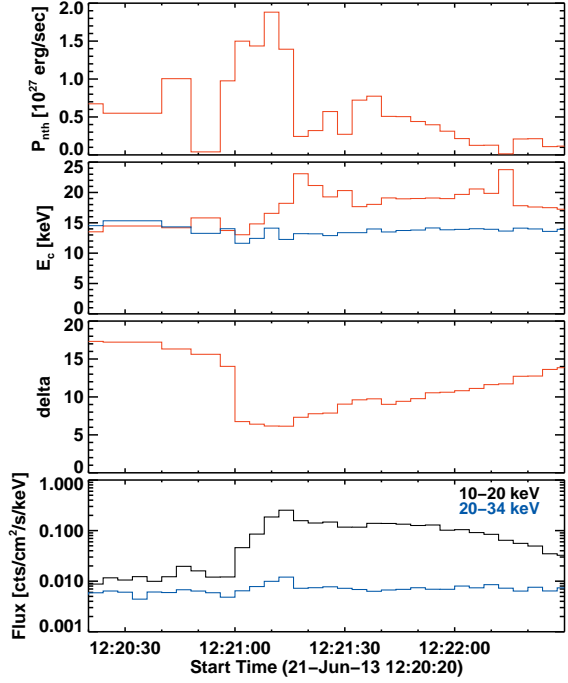}
\end{center}
\vspace{-1.0 cm}
\caption{Time evolution of the calculated main parameters of the thick-target model of the pre-impulsive and impulsive phases of the 21 June 2013 flare. From top to bottom: energy flux of non-thermal electrons $\rm P_{nth}$; calculated low cut-off energies $\rm E_{c}$ (red line) and low cut-off energies $\rm E_{c}$ derived from non-thermal component of the \emph{RHESSI} spectra (blue line); electron spectra index $\rm \delta$; \emph{RHESSI} fluxes in 10--20~keV (black line) and 20--34~keV (blue line) energy bands. The time resolutions of the $\rm P_{nth}$, $\rm E_{c}$ and $\rm \delta$ curves are 4~s during the impulsive phase but degrade to $\sim$20~s at the beginning of the flare. The time resolution of the \emph{RHESSI} fluxes is 4~s.
\label{Fig05}}
\end{figure}

Figure~\ref{Fig05} shows time variations of the main parameters of the NTEs beams powering the flare during the pre-impulsive and impulsive phases of the flare. The parameters were derived using the thick target model (version~2) \citep[see][for extended discussion]{Falewicz2014}. The flux of the non-thermal energy $\rm P_{nth}$ was calculated using the formula proposed by \citet{Tandberg1988}. During the pre-impulsive and impulsive phases of the flare the $\rm P_{nth}$ varied between $\rm P_{nth}=1.7\times10^{24}~erg\,s^{-1}$ and $\rm P_{nth}=1.9{\times}10^{27}~erg\,s^{-1}$. It reached the first local maximum at 12:20:40~UT, the second at 12:21:08 UT, and the third at 12:21:36~UT. The highest local maximum at 12:21:08~UT occurred during the impulsive phase of the flare. A strong increase of the $\rm P_{nth}$ before the second maximum was accompanied by an abrupt decrease of the electron spectral index $\delta$. At the same moment, the energy spectra of the NTEs hardened in accordance with the \textit{soft-hard-soft} pattern presented by the majority of solar flares during their impulsive phases \citep[see \emph{e.g.}][]{Grigis2004}. The first and second local maxima of the $\rm P_{nth}$ were also correlated with local maxima of the light curves recorded by {\it RHESSI} in the 10--20 keV and 20--34~keV energy ranges. The calculated values of the low energy cut-off $\rm E_c$ varied during the investigated period of the flare between 13~keV and 24~keV. These variations are introduced by the algorithm applied, which forces the observed and synthetic {\it RHESSI} 6--10~keV fluxes to be equal by changing the absorbed energy amount. The amplitude and time scale of the variations of the calculated $\rm E_c$ agree with the estimated ranges and time scales of the $\rm E_c$ variations which have been given by various authors for flares with higher GOES-classes \citep[see \emph{e.g.}][]{Holman2003,Sui2007,Hannah2008,Warmuth2009a,Warmuth2009b,Falewicz2011}.

\begin{figure}[t]
\figurenum{6}
\begin{center}
\includegraphics[width=7.5cm]{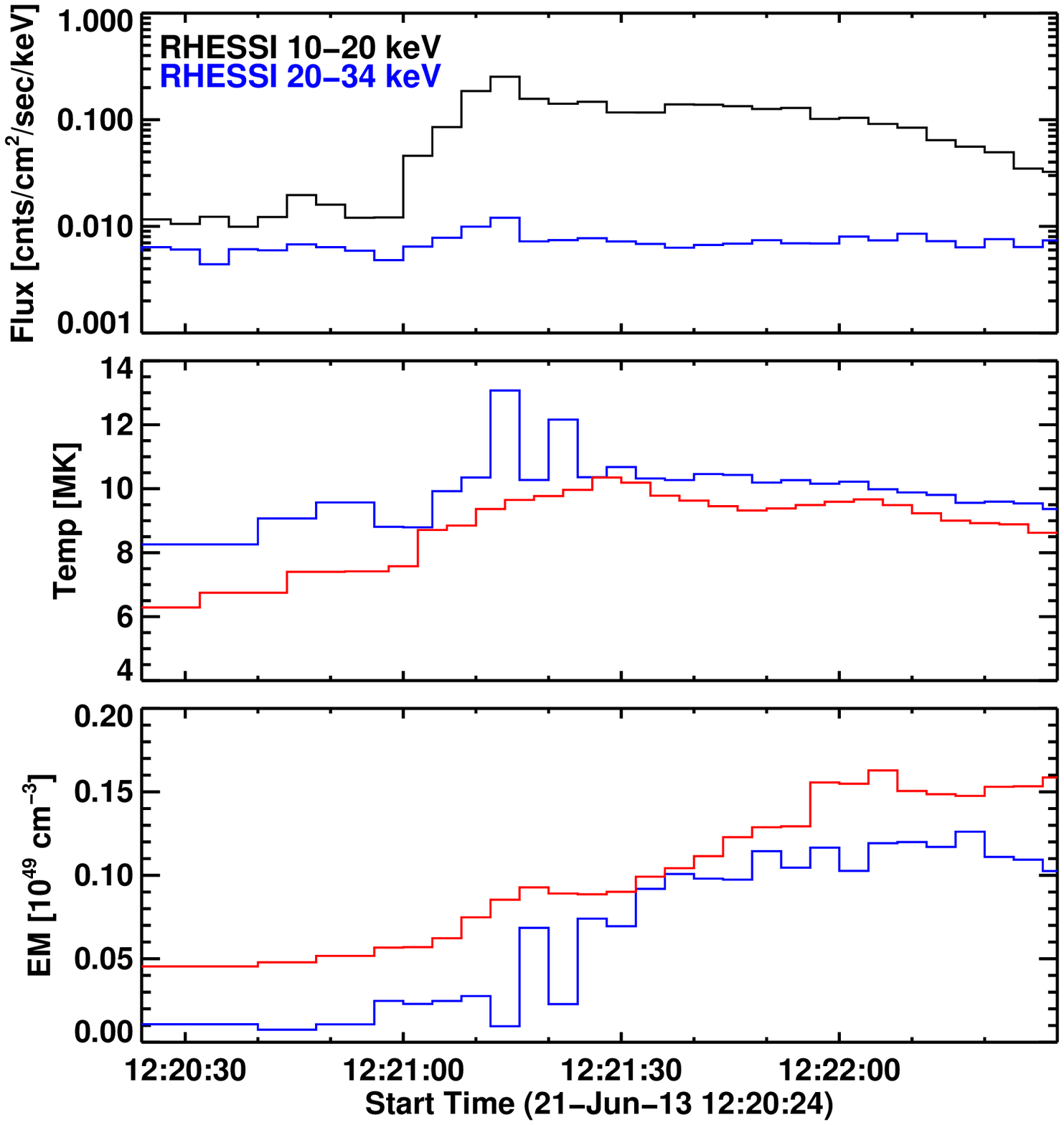}
\includegraphics[width=7.5cm]{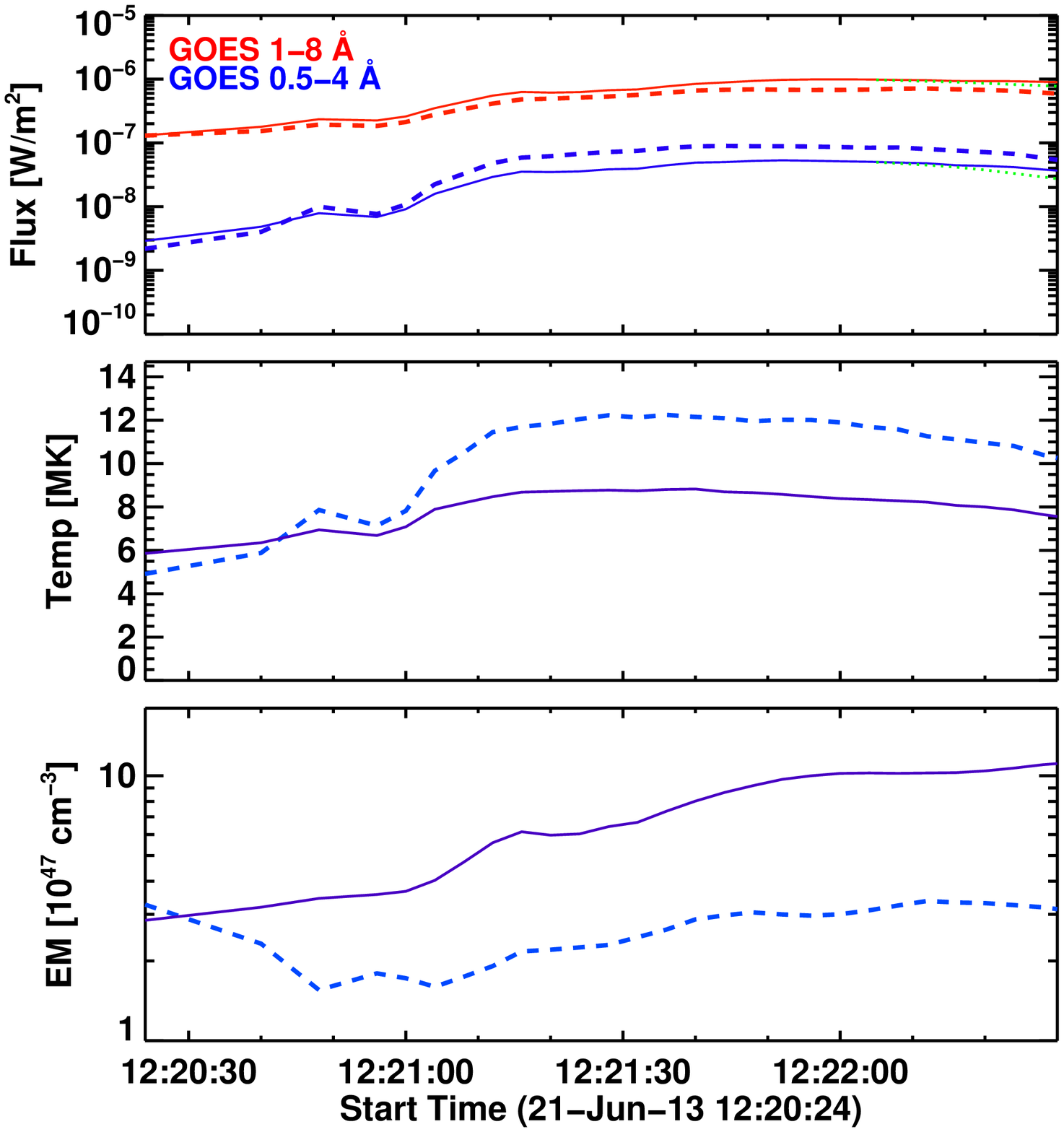}
\end{center}
\vspace{-1.5 cm}
\caption{\small Temperatures and emission measures derived from \emph{RHESSI} and \emph{GOES-15} data recorded before and during the impulsive phase of the 21 June 2013 flare compared with temperatures and emission measures calculated using the numerical 1D-HD model of the flare. Left column, from top to bottom: X-ray fluxes recorded by the {\it RHESSI} in the 10--20~keV and 20--34~keV energy bands; temperatures derived from the {\it RHESSI} spectra (blue line) and temperatures computed using the numerical model (red line); emission measures derived from the {\it RHESSI} spectra (blue line) and emission measures computed from the numerical model (red line). Right column, from top to bottom: X-ray fluxes recorded by {\it GOES--15} in the 0.5--4~{\AA} and 1--8~{\AA} energy bands (red and blue dashed lines) and {\it GOES} fluxes synthesized using the numerical model (red and blue solid lines); temperatures derived form {\it GOES} data (dashed light blue line) and {\it GOES} temperatures synthesized using the numerical model (solid dark blue line); emission measures derived from {\it GOES} data (dashed light blue line) and {\it GOES} emission measures synthesized using the numerical model (solid dark blue line). See main text for details.
\label{Fig06}}
\end{figure}

The temperatures and emission measures calculated from the {\it RHESSI} spectra and those representative of the numerical 1D-HD model of the flare are given in Figure~\ref{Fig06}, left column. The temperatures and emission measures calculated from the observed {\it GOES} fluxes and those computed using the numerical model are given in Figure~\ref{Fig06}, right column. The temperatures and emission measures were derived using the spectral fit of the thermal component of the \emph{RHESSI} spectra. The temperatures representative of the numerical model and given in Figure~\ref{Fig06} are equal to the temperatures of the plasma having the highest differential emission measure. The emission measures were calculated as
integrals of the computed differential emission measures for plasma having the temperatures greater than 1~MK. Before the maximum of the impulsive phase the temperatures of the plasma derived from {\it RHESSI} spectra were systematically higher than the temperatures calculated using the numerical model for the period. At the beginning of the pre-impulsive phase the differences between both temperatures were of the order of 2~MK, but later they  decreased gradually to less than 1~MK and completely vanished during the impulsive phase at 12:21:30~UT. At 12:21:12~UT and 12:21:20~UT the difference temporarily increased to about 4~MK and 3~MK, respectively, due to momentary increases of temperature derived from {\it RHESSI} spectra for the respective accumulation periods. The emission measures derived from {\it RHESSI} spectra were systematically lower than the synthesized ones by a factor not greater than 3--4, but during the maximum of the impulsive phase, at about 12:21:30~UT, the derived and synthesized emission measures were nearly the same.

\begin{figure}
\figurenum{7}
\begin{center}
\includegraphics[width=9.0cm]{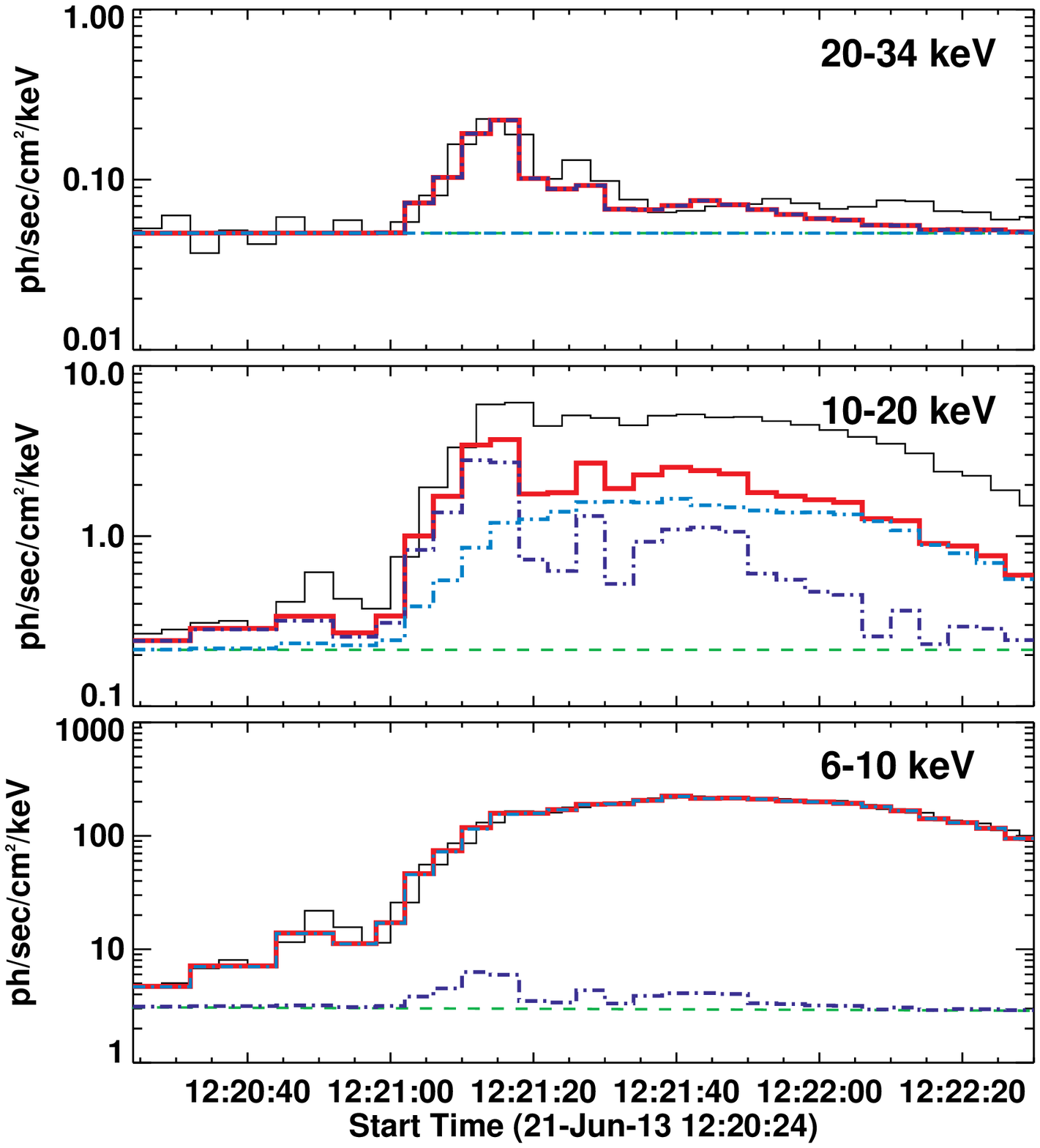}
\end{center}
\vspace{-1.0 cm}
\caption{\small {\it RHESSI} X-ray fluxes in the 6--10~keV, 10--20~keV, and 20--34~keV energy bands observed before and during the impulsive phase of the 21 June 2013 flare compared with the synthesized X-ray fluxes calculated using the numerical 1D-HD model of the flare. The observed X-ray fluxes are shown as black lines. The synthesized non-thermal and thermal fluxes are shown as dark blue and light blue dotted lines respectively. Red lines show the sums of the synthesized thermal and non-thermal fluxes. Dashed green lines show backgrounds derived from the observed data.
\label{Fig07}}
\end{figure}

The synthetic {\it GOES-15} 0.5-4~{\AA} and 1--8~{\AA} fluxes, calculated from the numerical model, follow qualitatively the observed fluxes, though the synthesized 1--8~{\AA} flux was always slightly higher than the observed flux. The synthesized 0.5--4~{\AA} flux was slightly higher than the observed one at the beginning of the flare only, but after 12:21:45~UT it became a factor of 2--3  lower. At the flare beginning, the temperatures calculated from {\it GOES-15} data and synthesized {\it GOES} temperatures were similar. However, after 12:21:00~UT the modeled temperatures were always lower than the observed values, at 12:21:20~UT the difference reached $\sim$2.5~MK. We believe this to arise through the variations of the 0.5--4~{\AA} to 1--8~{\AA} flux ratio which causes differences in estimated temperature and emission measure \citep{White2005}. The calculated emission measures were systematically greater than the emission measures derived from observations by a factor of about four.

The synthesized and observed total fluxes measured in the 6--10~keV, 10--20~keV and 20--34~keV energy bands are shown in Figure~\ref{Fig07}. The observed and synthesized light curves agree well during the flare impulsive phase. In the 20--34~keV energy band the emission is mostly caused by non-thermal processes and thus the time-variations of the non-thermal component define the time variations of the total emission in this energy band. The relative contribution of the non-thermal component decreases in the lower energy bands, where the thermal component plays a dominant role. The agreement of the observed and synthesized light curves in the 10--20~keV energy band is worse than for the higher and lower energy ranges, but the overall time course of the synthesized light curve, up to the impulsive phase of the flare, mimics qualitatively the observed one but is below the observed light curve. In the lowest energy band of 6--10~keV the thermal component dominates, although during the impulsive phase of the flare the non-thermal component also adds a small contribution to the total flux. The course of the synthesized emission in the 6--10~keV energy band is identical with the observed emission as a result of the applied method of the $E_c$ fitting.

\begin{figure}[t]
\figurenum{8}
\begin{center}
\includegraphics[width=9.0cm, height=6.0cm]{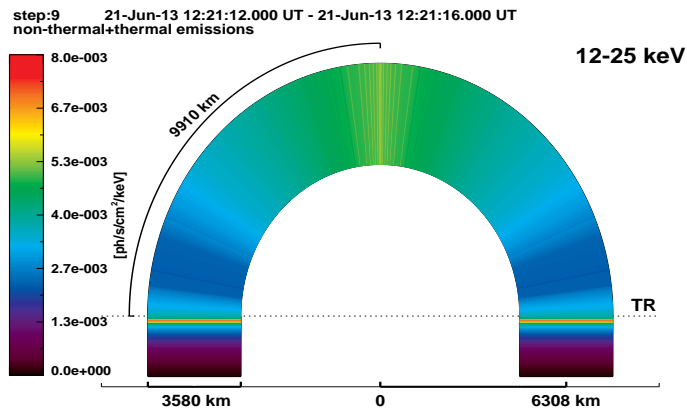}
\end{center}
\vspace{-1.0 cm}
\caption{\small Modeled distribution of the 12--25~keV X-ray emission along the loop during the impulsive phase of the 21 June 2013 flare. The dashed horizontal line indicates the transition region. Red layers below the transition region mark the impulsive brightenings at the feet of the loop.
\label{Fig08}}
\end{figure}

An example of the calculated spatial distribution of the X-ray emission along the flaring loop during the impulsive phase of the flare is given in Figure~\ref{Fig08}. Strong impulsive brightenings at the loop footpoints, and a much weaker and less well-defined non-thermal emission source at the top of the loop are evident. After 12:21:18~UT the emission of the flare was dominated by the thermal emission (see Figure~\ref{Fig07}, middle panel), and the top of the loop was the brightest in the 12--25~keV energy band (see Figure~\ref{Fig04}).

\begin{figure}
\figurenum{9}
\begin{center}
\includegraphics[width=6.5 cm]{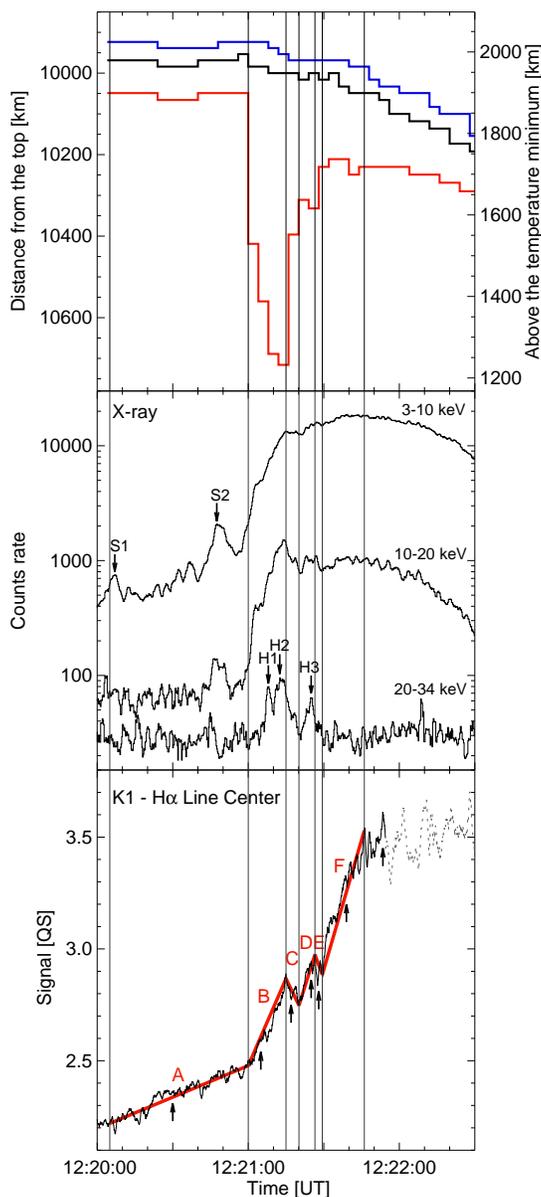}
\end{center}
\vspace{-1.0 cm}
\caption{\footnotesize Time variations of the calculated position and extent of the NTEs energy deposition region, the observed X-ray emission in various energy ranges, and the observed H$\alpha$ line centre (H$\alpha_0$) emission of the K1 flaring kernel during the 21 June 2013 flare. Upper panel: the position of the maximum of the NTEs deposited energy flux (black line), with lower and upper boundaries of the energy deposition region (red and blue lines, respectively). The lower boundary of the energy deposition region was arbitrarily selected at an energy flux of 0.01~erg/s/cm$^3$. Middle panel: X-ray emission (counts s$^{-1}$) in the \emph{RHESSI}  3--10~keV, 10--20~keV, and 20--34~keV energy ranges. The time resolution is 0.25~s after de-modulation of the original data; the data are smoothed with a 1~s wide box-car filter. The symbols S1, S2, H1, H2, and H3 indicate soft and hard X-ray impulses (see text). Lower panel: the  H$\alpha_0$ light curve of the K1 kernel. Black line: data with $\Delta$T=0.05~s time resolution, and smoothed using a 1~s wide box-car filter. Thick red lines indicate the main stages, marked from A to F (see text). Vertical lines delimit consecutive evolution stages of the overall variations of the H$\alpha$ line center light curve of the K1 kernel. The vertical arrows indicate the relevant moments in time given in Fig.~\ref{Fig11}. The H$\alpha$ signal is scaled to the signal of the quiet Sun (QS). The light curve is given in a wavelength reference system co-moving with plasma of the K1 kernel.
\label{Fig09}}
\end{figure}

The energy deposition rates were calculated using the approximative formula given by \citet{Fisher1989}. The NTEs deposited most of the energy inside a layer of limited width, called hereafter the energy deposition layer (EDL). The upper boundary of the EDL was defined as a plasma layer where the deposited energy flux starts to increase rapidly nearby the chromosphere; the lower boundary of the EDL was selected arbitrarily at an energy flux level of 0.01 erg/s/cm$^3$ (see Fig.~\ref{Fig09}, upper panel, where the time variations of the H$\alpha_0$ light curves are compared with X-ray light curves in three energy bands).

During the early phase of the flare (before 12:20:40~UT), the {\it RHESSI} X-ray fluxes in the 10--20~keV and 20--34~keV energy bands remained almost constant (Fig.~\ref{Fig09}, middle panel), and the maximum flux of the deposited energy was only $\sim10~erg/s/cm^{3}$, causing gradual heating of the local plasma. \emph{RHESSI} observed a pre-flare X-ray increase in the 3--10~keV energy band at about 12:20:25~UT. A well defined X-ray impulse was recorded at the end of the pre-impulsive phase at about 12:20:45~UT in the 20--34~keV band, and a little later in the lower energy bands. Due to the gentle evaporation of hot plasma caused by an ongoing low-intensity heating, the EDL maintained nearly constant thickness of about 125~km up to 12:21:00~UT, when it extended from $\rm L=9\,925$~km up to $\rm L=10\,050$~km, if counted from the top of the loop, or equivalently it extended between the altitudes $H=1900$~km and $H=2025$~km above the temperature minimum (Fig.~\ref{Fig09}, upper panel).

The impulsive phase of the flare started at about 12:21:00~UT in the 20--34~keV range and a few seconds earlier in lower energy bands. At the same time the EDL substantially widened, increasing its width to 725~km at 12:21:12 UT. The maximum flux of the deposited energy surged to $200-300$~erg/s/cm$^{3}$. Simultaneously, the upper boundary of the EDL fell to an altitude of $H=1975$~km above the temperature minimum due to increased plasma evaporation (see Fig.~\ref{Fig09}, upper panel). Three well distinguishable brief increases of the 20--34~keV X-ray flux were observed during the impulsive phase of the flare, at 12:21:08~UT, at 12:21:13~UT, and at 12:21:25~UT, respectively. These impulses are marked H1, H2 and H3 in Figure~\ref{Fig09}, middle panel. The second and third impulses were also noticeable in the 3--10~keV and 10--20~keV energy bands. The EDL shrank to roughly $D=425$~km at 12:21:17~UT after the H2 impulse, and to $D=250$~km at 12:21:28~UT after the H3 impulse. The EDL then gradually became even narrower, and simultaneously the X-ray emission substantially faded in all energy bands. Between 12:21:00~UT and 12:22:20~UT the upper boundary of the EDL persistently lowered from H=2025~km to H=1900~km above the temperature minimum due to the induced plasma evaporation. The maximum of the deposited energy flux moved in accordance with the upper boundary of the EDL. However, the abrupt increase of the extent of the EDL during the impulsive phase of the flare was caused mostly by changes of the position of the lower boundary due to an increased flux of the high energy NTEs precipitating further along the loop from the near loop-top release region to the dense plasma above the temperature minimum.

\section{Time Variations of the Observed H$\alpha$ Emission} \label{sec:halpha}

Two K1 and K2 H$\alpha$ flaring kernels were the brightest and the most active flaring structures visible in the H$\alpha$ light in the observed active region during the whole lifespan of the flare. The kernels were located close to both ends of the extended loop-like structure visible in the X-ray images. Between 12:20:00~UT and 12:21:55~UT the seeing conditions in the Bia{\l}k{\'o}w Observatory were very good, but later the clouds disturbed the observations. Thus, the variations of the H$\alpha$ emission after 12:21:55~UT are not investigated.

\begin{figure}[ht!]
\figurenum{10}
\begin{center}
\includegraphics[width=12.0cm]{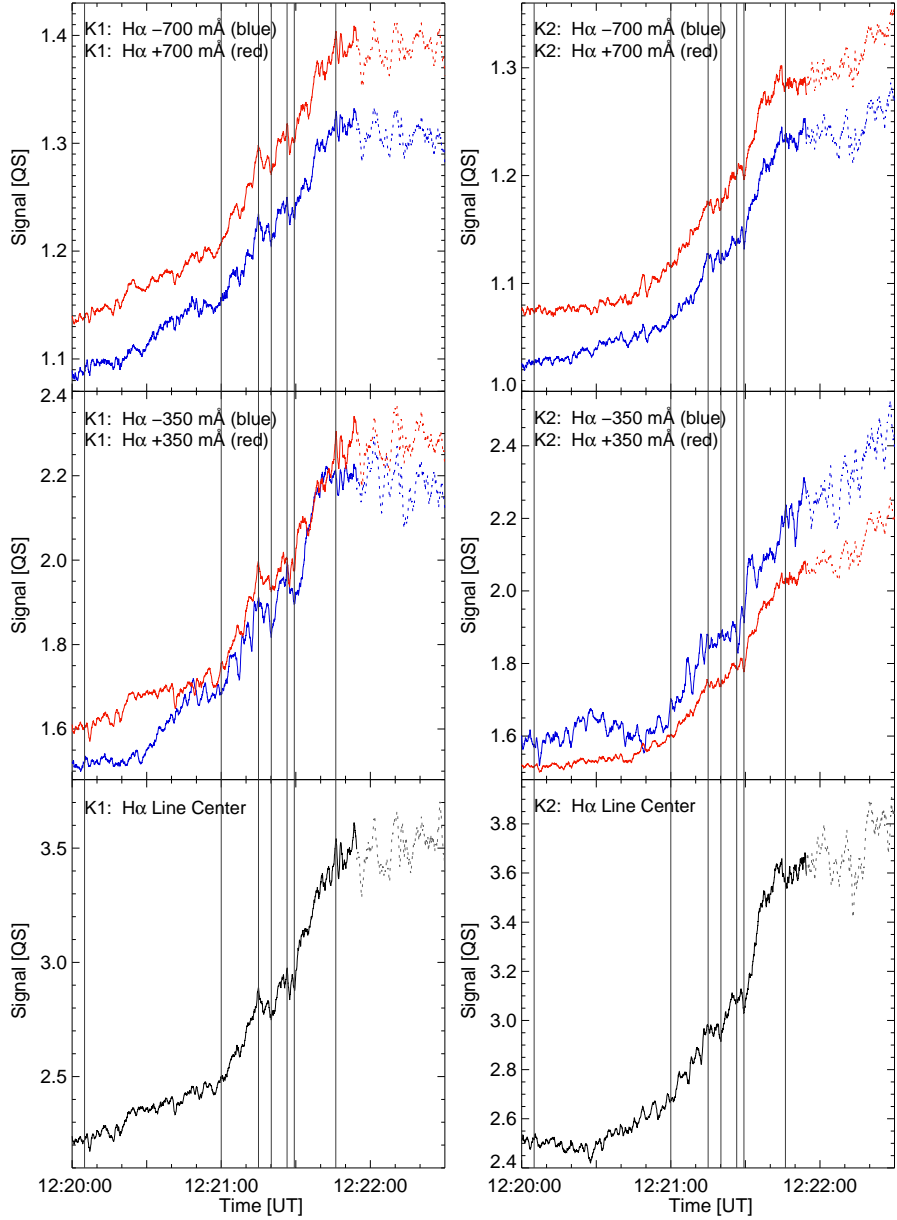}
\end{center}
\vspace{-1.0 cm}
\caption{\small H$\alpha$ line light curves of the flaring kernels K1 and K2 observed in the 21 June 2013 flare. The light curves are shown for the H$\alpha$ line center and for wavelengths $\Delta\lambda=\pm0.35~{\AA}$ and $\Delta\lambda=\pm0.70~{\AA}$ from the H$\alpha$ line center; the widths of both wavelength bands are equal to 0.06~\AA. The signals are scaled to the signal of the quiet Sun (QS). The light curves are shown in a wavelength reference system that is co-moving with plasma of each kernel. The raw data were collected with $\Delta$T=0.05~s time resolution, while the light curves shown are smoothed using a 1 s wide box-car filter. Vertical gray lines indicate the same moments as in Fig.~\ref{Fig09}. Data collected after 12:21:55~UT, presented as dotted lines, are disrupted by clouds.
\label{Fig10}}
\end{figure}

The K1 kernel started to brighten in the H$\alpha_0$, H$\alpha_0\pm0.35$~{\AA}, and H$\alpha_0\pm0.7$~{\AA} wavelengths immediately after the soft X-ray impulse observed in the 3--10~keV energy band at 12:20:07~UT (the impulse is marked S1 in Figure~\ref{Fig09}, middle panel), at the beginning of the pre-flare phase. The main stages of the emission of the K1 kernel in the H$\alpha_0$ wavelength are marked with thick red lines and with capital letters from A to F in the lower panel of the Figure~\ref{Fig09}. The emission of the K1 kernel in the blue wing of the H$\alpha$ line at H$\alpha_0$-0.35~{\AA} started to increase 20 seconds later, at about 12:20:30~UT (see Fig.~\ref{Fig10}, left column, middle panel). The light curves of the K1 and K2 kernels shown in Fig.~\ref{Fig09} and Fig.~\ref{Fig10} are scaled to the signal of the quiet Sun (QS), and the wavelengths of the emission are measured in the reference systems of the emitting plasma, so the influence of the solar rotation and plasma motions along the flaring loop is removed. The emission of the K1 in the line center H$\alpha_0$ during the pre-flare phase of the flare is marked by A. Immediately after the beginning of the impulsive phase, the K1 kernel brightened very rapidly in the H$\alpha_0$ band (phase B). This stage of evolution of the H$\alpha_0$ light curve of the K1 kernel ended abruptly at 12:21:15~UT, just after the X-ray impulse H2. The time delay between the maximum of the H2 impulse, recorded at 12:21:13~UT in the 20-34~keV energy band, and the corresponding maximum of the H$\alpha_0$ light curves of the K1 kernel was equal to only 2~s in all five  wavelengths: H$\alpha_0$, H$\alpha_0\pm0.35$~\AA, and H$\alpha_0\pm0.7$~{\AA} (Fig.~\ref{Fig10}, left panel). Later, up to 12:21:20~UT, the H$\alpha_0$ emission of the K1 faded gradually in accordance with the decreasing X-ray fluxes recorded in all investigated energy bands (stage C). At 12:21:20~UT the emission of the K1 started to increase again, simultaneously with the beginning of the H3 X-ray impulse recorded in the 20--34~keV energy band (stage D). The growth of the emission stopped at 12:21:27~UT, simultaneously with the end of the H3 impulse. A brief subsequent decline of the emission of the K1 (stage E) occurred between 12:21:27~UT and 12:21:29~UT, when the X-ray flux in the 20-34~keV remained stable, but the fluxes in the 3--10~keV (mostly thermal) and 10--20~keV bands slightly decreased. The subsequent growth of the K1 emission in the H$\alpha_0$ light (stage F) occurred during the post-impulsive phase of the flare, when the X-ray emission stabilized in the 10--20 keV and 20--34~keV energy bands, but the emission in the 3--10~keV energy band still rose due to the increasing volume of the hot evaporated plasma. The emission of the K2 kernel in the H$\alpha_0$ wavelength increased at 12:20:30~UT, roughly 20~s after the start of the emission of the K1 kernel (Fig.~\ref{Fig10}, right panel). The time variations of the emission of the K2 kernel were similar to the variations of the K1 kernel. However, during stage C the emission of the kernel K2 remained nearly constant in the H$\alpha_0$ wavelength.

\begin{figure}[t]
\figurenum{11}
\begin{center}
\includegraphics[width=14.0cm]{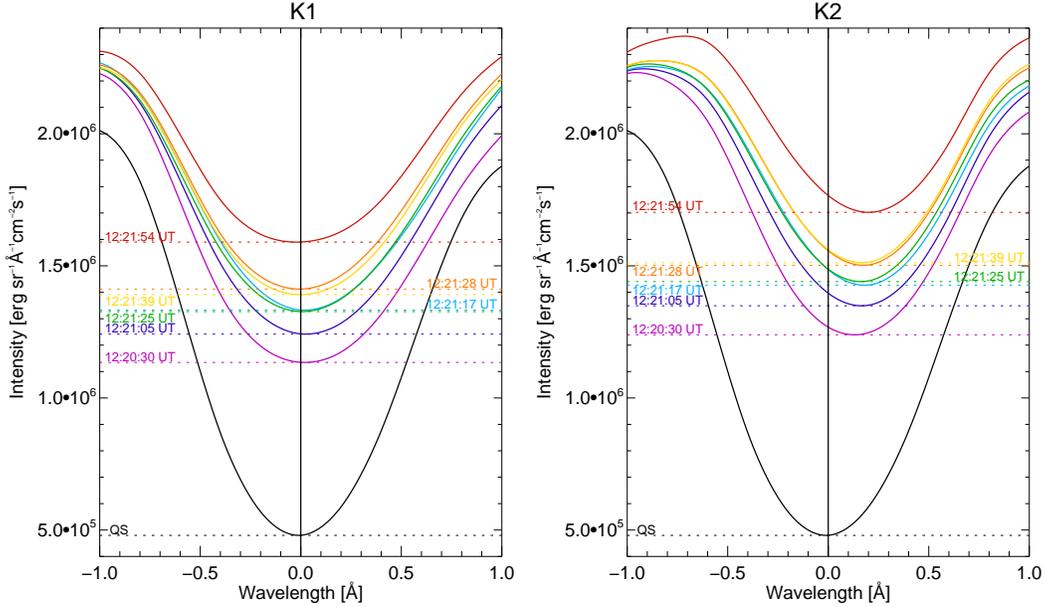}
\end{center}
\vspace{-1.0 cm}
\caption{\small Averaged H$\alpha$ line profiles emitted by the flaring kernels K1 (left panel) and K2 (right panel) during the 21 June 2013 solar flare at 12:20:30~UT (violet line), 12:21:05~UT (dark blue), 12:21:17~UT (blue), 12:21:25~UT (green), 12:21:28~UT (orange), 12:21:39~UT (yellow), and 12:21:54~UT (red). The relevant emission intensities in the H$\alpha$ line center are marked with horizontal dotted lines with the same colors. The same times are indicated by the vertical arrows in Fig.~\ref{Fig09}. The time resolution of the spectral observations was 0.05~s. The bandwidth of the profiles is limited to $\pm1.0~{\AA}$ from the H$\alpha$ line center. The mean H$\alpha$ line profile of the nearby quiet solar chromosphere (QS) is also shown (black line).
\label{Fig11}}
\end{figure}

\begin{figure}[h!]
\figurenum{12}
\begin{center}
\includegraphics[width=7.0cm]{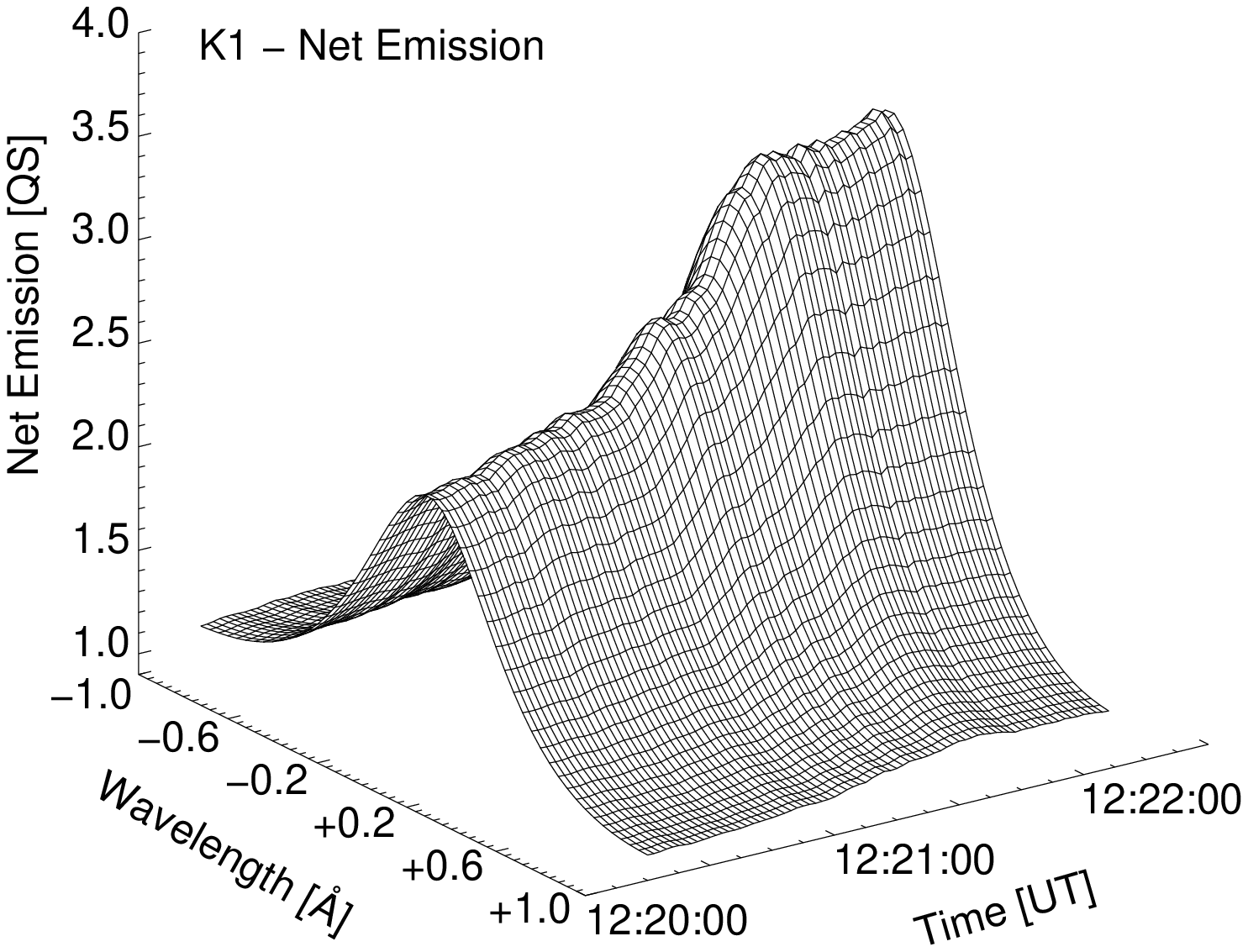}
\includegraphics[width=7.0cm]{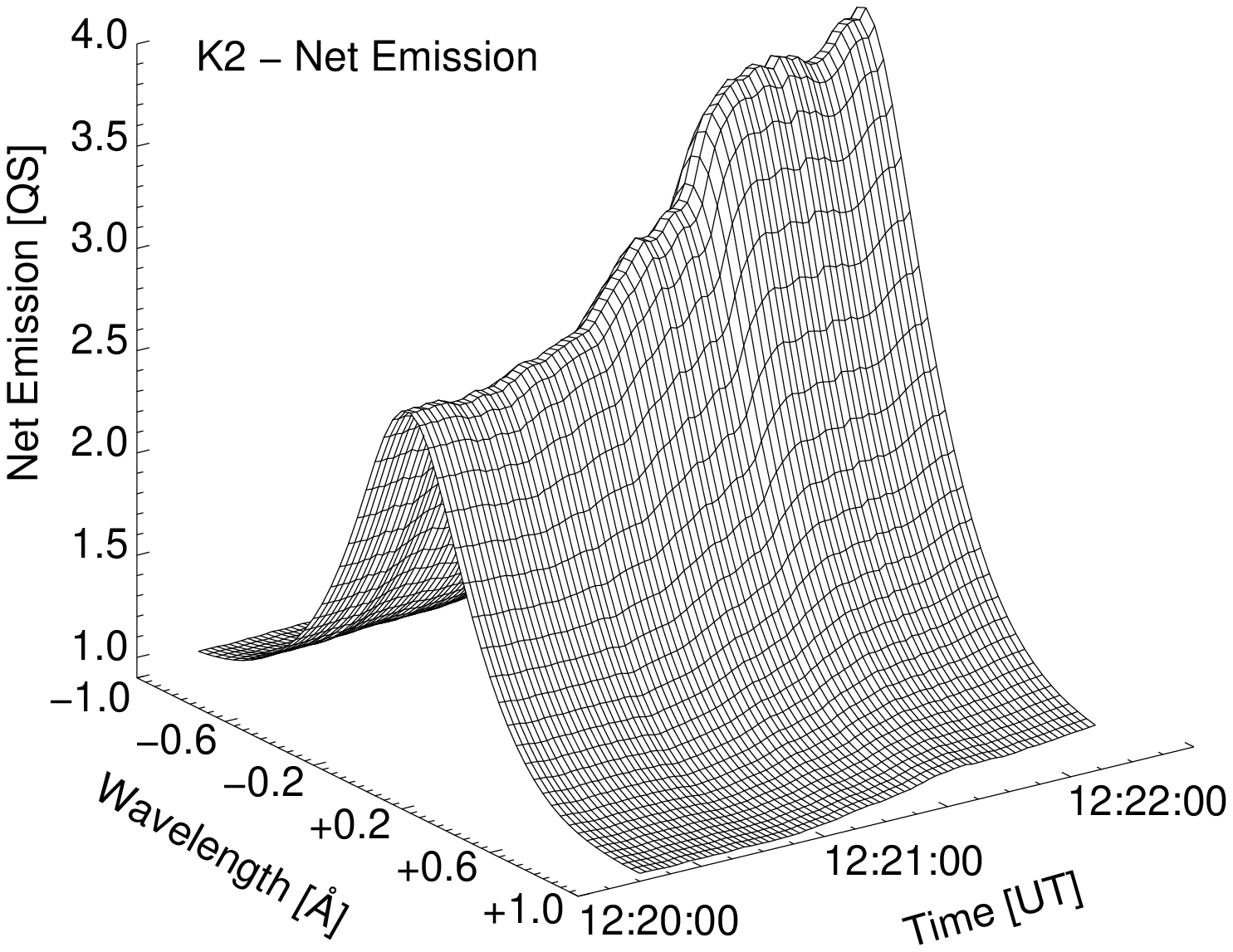}
\end{center}
\vspace{-1.0 cm}
\caption{\small Averaged H$\alpha$ net emission from the flaring kernels K1 and K2 during the 21 June 2013 solar flare between 12:20:00~UT and 12:22:30~UT.
\label{Fig12}}
\end{figure}

The time variations of the emissions of the K1 and K2 kernels in opposite line wings were different. At $\Delta\lambda=\pm0.7$~{\AA} from the H$\alpha$ line center, the emission of both kernels was stronger in the red wing than in the blue wing. However, at $\Delta\lambda=\pm0.35$~{\AA} from the line center the K2 kernel was brighter in the blue wing, but the K1 was slightly brighter in the red wing (Fig.~\ref{Fig10}). During the entire period the profile of the H$\alpha$ line emitted by the K2 kernel was persistently red-shifted, while the H$\alpha$ line emitted by the K1 kernel did not show any significant shift. The mean H$\alpha$ line profiles of the kernels K1 and K2 are shown in Figure~\ref{Fig11} for seven selected moments showing main stages of the evolution of the K1 kernel. The same moments are also marked by seven vertical arrows in Figure~\ref{Fig09}. The net H$\alpha$ emission of the kernels K1 and K2 between 12:20:00~UT and 12:24:00~UT (defined as the kernel emission with emission of the quiet chromosphere subtracted) are given in Figure~\ref{Fig12}. The shapes and Doppler shifts of the H$\alpha$ line varied after the HXR impulses, e.g. at 12:21:09~UT, one second after the H1 peak of the HXR emission at 12:21:08~UT.

\begin{figure}[h!]
\figurenum{13}
\begin{center}
\includegraphics[width=8.0cm]{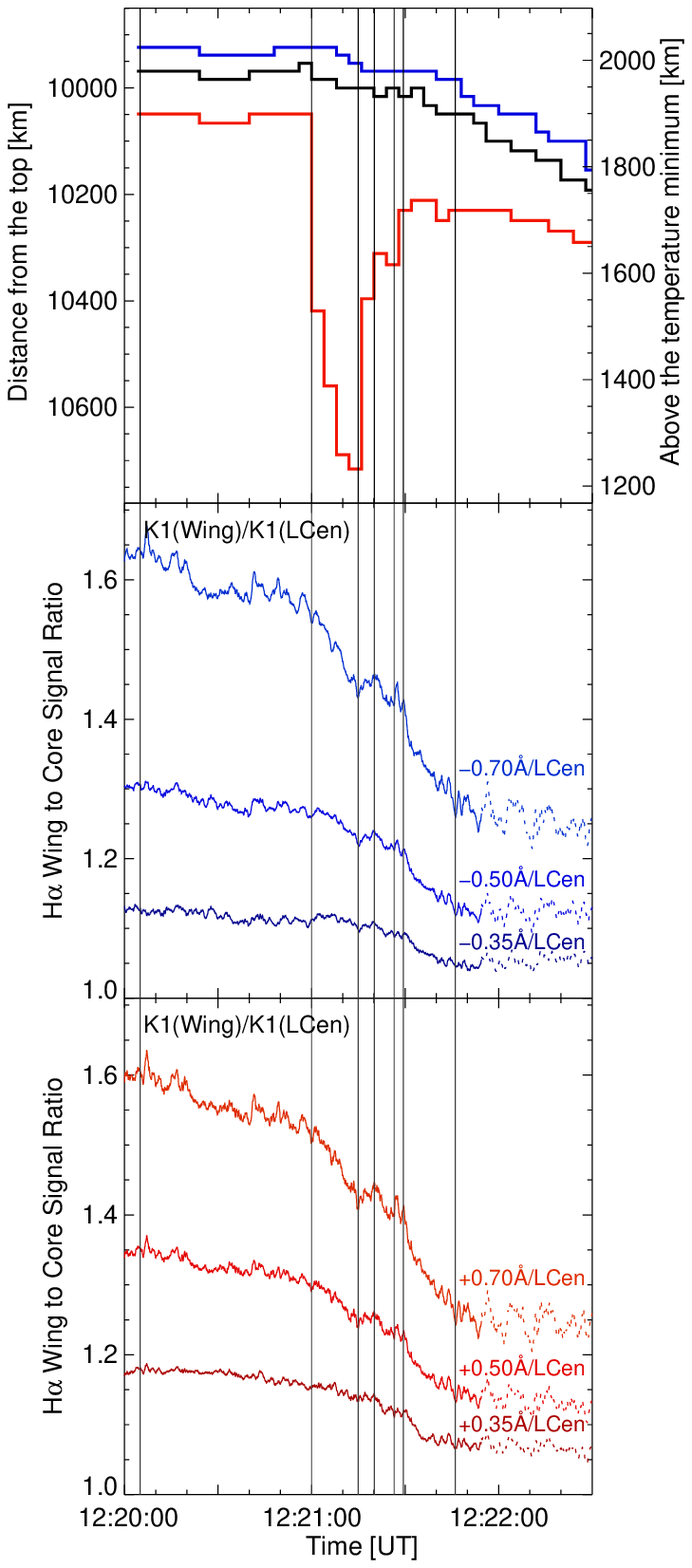}
\includegraphics[width=8.0cm]{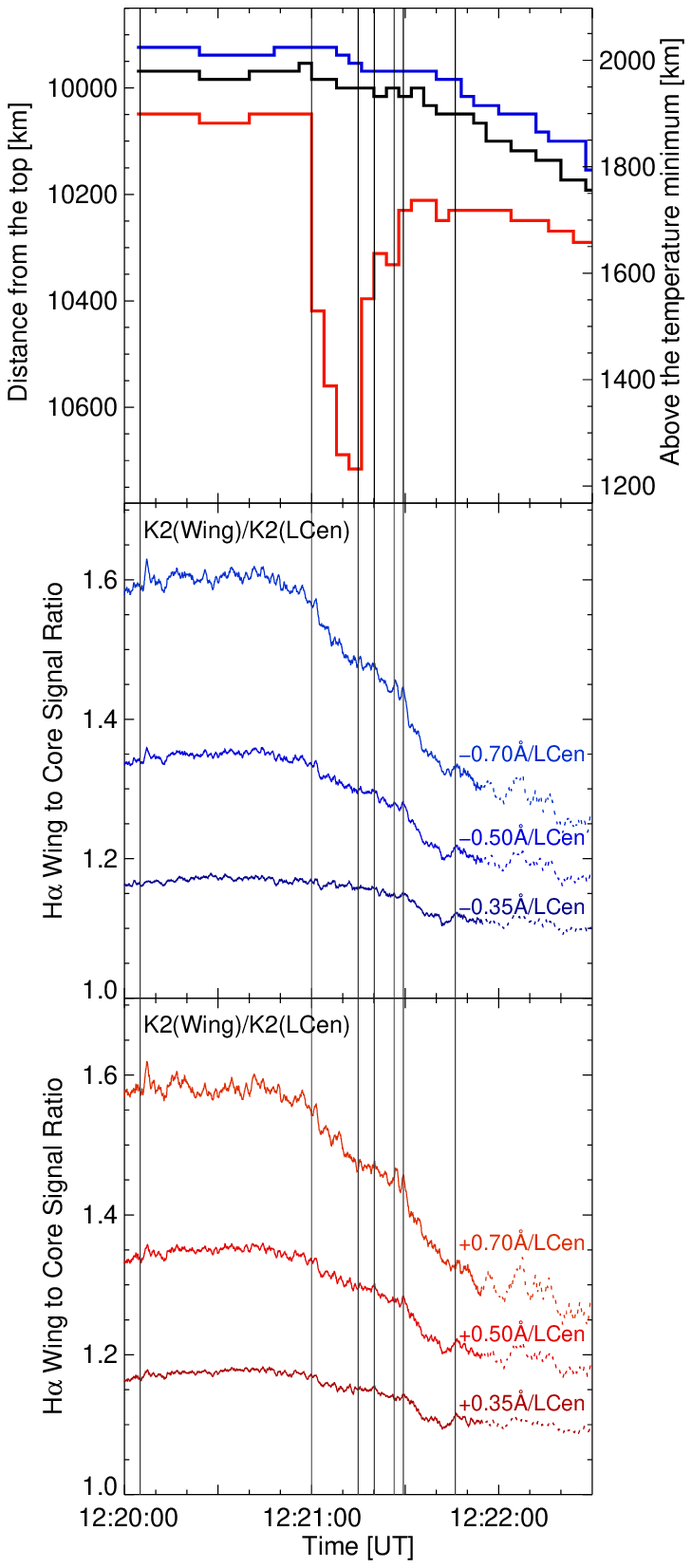}
\end{center}
\vspace{-1.0 cm}
\caption{\small Time variations of the wing-to-line center (i.e. wing to core) intensity ratios of the flaring kernels K1 and K2 for wavelengths
$\Delta\lambda=\pm0.35~{\AA}$, $\Delta\lambda=\pm0.5~{\AA}$, and $\Delta\lambda=\pm0.7~{\AA}$ from the H$\alpha$ line center (all wavelengths are corrected on Doppler effect, i.e. corrected for bulk plasma motion, being measured from the FWHM center). Vertical gray lines indicate the same moments as in Fig.~\ref{Fig09}. Upper panel is the same as in Fig.~\ref{Fig09}.
\label{Fig13}}
\end{figure}

The variations of the wing-to-line center intensity ratios (IRs), particularly the IRs measured in the selected wavelengths in the line wings of the H$\alpha$ line at $\Delta\lambda=\pm0.35$~{\AA}, $\Delta\lambda=\pm0.5$~{\AA}, and $\Delta\lambda=\pm0.7$~{\AA} as well as in the line center H$\alpha_0$ are given in Figure~\ref{Fig13}, where the wavelengths are corrected for Doppler effect, i.e. shifts due to bulk plasma motions are removed. A slow, gradual decrease of the IRs at $\Delta\lambda=\pm0.7$~{\AA} occurred after the first soft X-ray impulse S1 recorded in the 3--10~keV energy band at 12:20:07 UT. After 12:21:18 UT (time of the H1 impulse in the 20--34~keV energy band), the IRs decreased abruptly. Short-lasting increases of the IRs were correlated in time with X-ray impulses recorded both in HXR and SXR ranges. The time delays between X-ray impulses and abrupt variations of the H$\alpha$ profiles varied between zero (the X-ray impulse at 12:21:13~UT and H$\alpha$ emission at $\Delta\lambda=\pm0.7$~\AA), and 2.1~s (the X-ray impulse at 12:21:25~UT and H$\alpha$ emission at $\Delta\lambda=\pm0.7$~\AA).

\section{Discussion and Conclusions} \label{sec:discus}

Results from a 1D-HD numerical model of the C1.1 {\it GOES}-class solar flare observed on 21 June 2013 in the NOAA~11772 active region are presented here. The model is calculated using geometric and thermodynamic parameters derived from \emph{\emph{RHESSI}} data and is powered by the energy deduced from \emph{RHESSI} X-ray spectra on the assumption that NTEs delivered external energy to the flaring loop. The \emph{RHESSI} spectra of the flare show a clear NT component, particularly during the impulsive phase, from which the NTEs energy flux can be derived. The low-energy cut-off energies of the electron beams were automatically optimized at each time step of the calculations by fitting synthetic flux to the flux recorded by {\it RHESSI} in the 6--10~keV energy band.

Substantial differences between the observed and the synthesized light curves are noticeable in the 10--20~keV energy range. An underestimation of the plasma's thermal emission can be attributed to two main factors. First, the spatial distribution of the local thermodynamic and kinematic parameters of the plasma in the numerical model do not strictly match the actual spatial distribution of the plasma inside the real flaring loop. Secondly, the assumed power-law energy spectrum of the injected NTEs also caused some differences between the modeled and observed X-ray fluxes in the 10--20~keV energy band. \citet{Liu2009} showed using the Fokker-Planck model of the particle transport, that if the energy spectrum of the injected NTEs is shaped by a stochastic acceleration, there appear higher coronal temperatures and densities, larger up-flow velocities, and faster increases than if the electron energy spectrum is of power-law type. This is because the energy spectrum of the injected electrons ranges smoothly from a quasi-thermal component to a non-thermal tail. The injected beam of electrons generally includes much energy in the form of low-energy electrons that deposit their energy primarily in the upper part of the loop, and indirectly enhances the evaporation. Thus, the thermal emission component could give a greater contribution to the total flux in the 10--20~keV energy range. In the higher energy range of 20--34~keV, which is dominated by the non-thermal component, the differences between the modeled and the observed fluxes are much smaller.

The properties and basic assumptions of the applied 1D-HD numerical model were discussed above. Inconsistencies between the modeled spatial distribution of the plasma parameters and the actual distribution of the plasma parameters in the flaring loop were obviously caused by simplifications of the applied 1D-HD numerical code, rough estimation of the geometric parameters of the loop, errors in the restoration of the {\it RHESSI} spectra, errors in the derivation of the NTEs beam parameters, and an application of a simplified analytical formula of the  heating function of the NTEs. The geometric characteristics of the modeled loop are estimated very inaccurately due to the limited spatial resolution of the \emph{RHESSI} instrument and due to the applied assumptions about the shape and position of the loop. Due to the small cross-section of the flaring loop and the limited spatial resolution of \emph{RHESSI}, the formal uncertainties on the cross-section of the modeled loop is 80\%. The uncertainty of the cross-section causes a proportional uncertainty of the local density of the delivered energy. The alternation of the local energy density substantially changes the resulting numerical model, e.g. by influencing the dynamics of plasma and altering the resulting class of the flare. The uncertainties of the fitting of the {\it RHESSI} spectra, including the applied model of the fit and the uncertainties of the observational data directly influence the estimated parameters of the NTE beams and their energy. Due to a complicated problem of the influence of the errors of the parameters on the resulting numerical model, the uncertainties of the model could only be estimated using the Monte Carlo method on the spectrum and other parameters, which will be evaluated in future work. The uncertainties introduced by the simplified analytical formula of the NTE heating function are less important. The differences between the heating function calculated using the approximate formula and the one calculated using Fokker-Planck method are minor in the lower part of the loop, but increase in the upper part of the loop, where the column mass is relatively low. Therefore, plasma in the lower part of the loop is powered by well approximated energy flux. It must also be considered that real flaring loops possess the multi-thread internal structure, which is poorly resolved by the contemporary observing instruments, and which is not taken into account in the applied 1D-HD numerical models. Similarly, there are no direct methods  to observe the heating episodes and the thermodynamic evolution of the individual threads. Despite all these limitations of the 1D-HD models, these models, when powered by the appropriate energy flux derived from the \emph{RHESSI} data, yields a good agreement of the predicted and observed X-ray fluxes during the impulsive phase of the flare, highlighting that these models sufficiently describe the plasma processes during the impulsive phase despite many simplifications, and show that the plasma heating during the impulsive phase is adequately powered by the NTEs.

The direct comparison of the calculated variations of the plasma characteristics in the numerical model of the flaring loop with the observed variations of the H$\alpha$ emission of the solar flare was possible only due to exceptional parameters of the applied spectrograph, that is capable to record the H$\alpha$ line spectra with the very high cadence of 0.05~sec, and for all pixels inside the FOV. This data allows one to reconstruct high-cadence series of the quasi-monochromatic images of the whole FOV and the light curves of the selected flaring kernels taken in various wavelengths. The time variations of the H$\alpha$ line profiles and the emission intensities in the selected wavelengths were compared with the spatial and the temporal variations of the main parameters of the 1D-HD numerical model, used particularly with time variations of the position and thickness of the EDL.

In accordance with our previous results \citep{Radziszewski2007,Radziszewski2011}, the time variations of the X-ray fluxes in all investigated energy bands show good correlation, both in time and space, with the time variations of the H$\alpha$ line intensities and H$\alpha$ line profiles emitted by the H$\alpha$ flaring kernels, particularly for the K1 kernel during the pre-impulsive and impulsive phases of the flare. The observed responses of the H$\alpha$ line to the variations of the NTE heating were instantaneous or nearly so. For example, the time delay between the maximum of the H2 impulse, recorded at 12:21:13~UT in the 20--34~keV energy band, and the relevant local maxima of the H$\alpha$ emission of the K1 kernel was only 2~s in all five wavelengths (H$\alpha_0$, H$\alpha_0\pm0.35$~{\AA}, and H$\alpha_0\pm0.7$~{\AA}), illustrating the fast response of the chromosphere to the abrupt heating by the beam of NTEs, and the immediate energy loss by radiation. The line cores and wings of the mean H$\alpha$ profiles of flaring kernel K1 and K2 exhibit fast increases of the emission and variations of the line shapes and shifts after 12:21~UT, during the impulsive phase of the flare, when the flux of the NTEs was highest and the EDL was most extended. The variable beams of the NTEs precipitated at that time through the plasma to the middle chromosphere, where the wings of the H$\alpha$ line are formed. Besides, NTEs collide also with atoms and they can additionally ionize species, leading to an enhancement of an electron density in deposition layers, and causing an additive enhancement of the emission in the chromospheric lines, including H$\alpha$ \citep{Berlicki2005}. After the impulsive phase of the flare, the enhanced emission in the core of the H$\alpha$ line may be caused also by the conduction of energy  \citep{Czaykowska2001}. The wing-to-center ratios of the mean H$\alpha$ profiles decreased between 12:20~UT and 12:22~UT due to an increased emission in the line core. Usually cores of the chromospheric lines are significantly enhanced in flaring kernels, while sometimes their emission becomes higher than in the nearby continuum. However, the flare was a low-class event and the core of the H$\alpha$ line was not particularly bright. The decrease of the IRs suggests that significant energy was deposited by NTEs in the upper chromosphere, which should happen in the case of a weak flare with a low number of high-energy NTEs \citep{Falewicz2009a,Reep2013}. The line profiles emitted by K1 kernel were generally symmetric during the flare, but some  small brief shifts were observed during the rapid variations of the HXR emission. The variations of the H$\alpha$ light curve of the K1 kernel were correlated in time with the variations of the HXR better than the same variations of the K2 kernel. The constant red-shift (up to about +0.2~\AA) of the H$\alpha$ line profiles emitted by the K2 flaring kernel suggests a macroscopic downward plasma motion.

The vertical extent and location of the EDL was derived from the numerical model under the assumption that the upper boundary of the EDL was located in the plasma layer where the deposited energy flux starts to increase rapidly near the chromosphere. The lower boundary of the EDL was selected arbitrarily at the altitude at which the energy flux fell to 0.01 erg/s/cm$^3$. Before the impulsive phase of the flare, the EDL maintained nearly constant thickness of about 125 km, and extended between the altitudes $H=1900$~km and $H=2025$~km above the temperature minimum. During the flare impulsive phase, the calculated altitude of the lower boundary of the EDL abruptly fell and the effective thickness of the EDL quickly increased to about 725~km. This increase was caused by an enhanced flux of the energy deposited by NTEs, which precipitated deeper along the loop toward the temperature minimum. After the impulsive phase, the calculated extent of the EDL gradually shrank due to the decreased flux of the delivered energy. At the same time the upper boundary of the EDL slowly declined due to continuing chromospheric evaporation.

Assuming a semi-circular loop geometry perpendicular to the solar surface, the apparent positions of the centers of gravity of a foot point observed in various energy bands can be converted into their positions along the loop. For the brighter, north-west foot point, during the impulsive phase of the flare the difference of altitudes was equal to $\Delta$h~=~2550~km in the case of the overlapped 15--26~keV and 23--34~keV energy bands. The numerically modeled thickness of the EDL during the impulsive phase of the flare was about 3.5 times smaller. However, the extent and altitude of the EDL in the 1D-HD model were derived on many assumptions, as well as the estimations of the relative positions of the foot observed in various energy bands were approximate only. The relatively low energy fluxes emitted by the flare prevented an application of images restored in the non-overlapping energy bands and the evaluation of the vertical extents before and after the maximum. The modeled variations of the size and location of the EDL yield interesting correlations with the H-alpha behavior during the impulsive phase, from which certain physical processes of energy deposition can be better understood. Furthermore, the comparison of the time evolution of the EDL position and thickness with the time variations of the H$\alpha$ emission reveals that the variations of the H$\alpha$ profiles and intensities are well correlated in time with the modeled variations of the plasma densities and temperatures in the footpoints of the flaring loop, and modeled variations of the energy deposition depth. Before the impulsive phase, the mean intensity of the emission of the K1 kernel in the H$\alpha_0$ wavelength slowly increased by about 11 percent, up to 2.45 times the intensity of the H$\alpha$ emission of the quiet Sun. Later, during the impulsive phase of the flare, the H$\alpha_0$ intensity increased to 2.85 times the intensity of the emission of the quiet Sun, i.e. by another 16 percent, and simultaneously the calculated lower boundary of the EDL abruptly fell and the calculated thickness of the EDL increased to 725 km. After the impulsive phase, the thickness of the EDL diminished again due to the decreased flux of the energy, and for the same reason, the H$\alpha_0$ emission decreased at the same time. Next the H$\alpha$ emission persistently increased again because the volume of the emitting plasma increased as a result of the plasma evaporation. The H$\alpha$ profiles were shifted to longer wavelengths at that time, as a result of the slow downward motion of the emitting plasma.

Our results are consistent with an assumption, that time variations of the H$\alpha$ line profiles emitted by flaring kernels are caused primarily by temporal variations of penetration depths of non-thermal electron beams along flaring loops. The result emphasizes the importance of very fast variations of plasma properties in the lower part of flaring loops, where the chromospheric emission appears. The short, sub-second variations of the plasma parameters, revealed as a result of the numerical modeling, are comparable with the time scales of the chromospheric emission variations. The high-time resolution spectral and photometric observations of the flaring kernels, easily achievable with the 2D imaging spectrographs, are of a great value for the understanding of fast processes in the flaring chromosphere, as well as for investigations of the coupling between NTE beams, HXR, and chromospheric emissions.

\section{Acknowledgments}
The authors acknowledge the anonymous Referee for extended discussions and numerous, very constructive remarks, which allowed us to significantly improve the work. The authors acknowledge also the \emph{RHESSI} and \emph{SDO} consortia for providing the excellent observational data. The numerical simulations were carried out using resources provided by the Wroc{\l}aw Centre for Networking and Supercomputing (http://wcss.pl), grant No. 330. The research leading to the obtained results has received the funding from the European Community's Seventh Framework Programme ([FP7/2007-2013]) under the grant agreement no. [606862] -- F-CHROMA Project. PR was supported by the National Science Centre, Poland, under the grant no. UMO-2015/17/B/ST9/02073. AB was supported by the project no. 16-18495S of the Grant Agency of the Czech Republic and by the institutional grant RVO 67985815.


\begin{thebibliography}{}

\bibitem[Abbett \& Hawley(1999)]{Abbett1999} Abbett, W. P., \& Hawley, S. L. 1999, \apj 521, 906
\bibitem[Allred et al.(2005)]{Allred2005} Allred, J. C., Hawley, S. L., Abbett, W. P., \& Carlsson, M. 2005, \apj 630, 573
\bibitem[Antonucci et al.(1984)]{Antonucci1984} Antonucci, E., Gabriel, A. H., \& Dennis, B.R. 1984, \apj, 287, 917
\bibitem[Antonucci et al.(1999)]{Antonucci1999} Antonucci, E., Alexander, D., Culhane, J. L., et al. 1999, in The Many Faces of the Sun: a Summary of the Results from NASAs Solar Maximum Mission, ed. K. T. Strong, J. L. R. Saba, B. M. Haisch, \& J. T. Schmelz (Springer-Verlag), chapt. 10.
\bibitem[Aschwanden et al.(1999)]{Aschwanden1999} Aschwanden, M. J., Fletcher, L., Sakao, T., Kosugi, T., \& Hudson, H. 1999, \apj, 517,977
\bibitem[Aschwanden(2005)]{Aschwanden2005} Aschwanden, M. J. 2005, Physics of the Solar Corona (Berlin: Springer Praxis Books)
\bibitem[Aschwanden et al.(2016)]{Aschwanden2016} Aschwanden, M. J.,  Holman, G., O'Flannagain, A., Caspi, A., McTiernan, J., \& Kontar, E. P. 2016, \apj, 832, 27
\bibitem[Bai \& Ramaty(1978)]{Bai1978} Bai, T., \& Ramaty, R. 1978, \apj, 219, 705
\bibitem[Battaglia et al.(2012)] {Battaglia2012} Battaglia, M., Kontar, E. P., Fletcher, L., \& MacKinnon, A. L. 2012, \apj, 752, 4
\bibitem[Berlicki(2007)]{Berlicki2007} Berlicki, A., ASP Conference Series, Vol. 368, 387
\bibitem[Berlicki \& Heinzel(2004)]{Berlicki2004} Berlicki, A. \& Heinzel, P. 2004, A\&A, 420, 319
\bibitem[Berlicki et al.(2005)]{Berlicki2005} Berlicki, A., Heinzel, P.,Schmieder, B., Mein, P., Mein, N. 2005, A\&A, 430, 679
\bibitem[Brown(1971)]{Brown1971} Brown, J.C. 1971, \solphys, 18, 489
\bibitem[Czaykowska et al.(2001)]{Czaykowska2001} Czaykowska, A., Alexander, D., \& de Pontieu, B. 2001, \apj, 552, 849
\bibitem[Dere et al.(1997)]{Dere1997} Dere, K. P., Landi, E., Mason, H. E., Monsignori-Fossi, B. C.,\& Young, P. R. 1997, A\&A, 125, 149
\bibitem[Falewicz et al.(2009a)]{Falewicz2009a} Falewicz, R., Rudawy, P., \& Siarkowski, M. 2009a, A\&A, 500, 901
\bibitem[Falewicz et al.(2009b)]{Falewicz2009b} Falewicz, R., Rudawy, P., \& Siarkowski, M., 2009b, A\&A, 508, 971
\bibitem[Falewicz et al.(2011)]{Falewicz2011} Falewicz, R., Siarkowski, M., \& Rudawy, P., 2011, \apj, 733, 37
\bibitem[Falewicz(2014)]{Falewicz2014} Falewicz, R. 2014, \apj, 789, 71
\bibitem[Falewicz et al.(2015)]{Falewicz2015} Falewicz, R., Rudawy, P., Murawski, K., Srivastava, A. K. 2015, \apj, 813, 70
\bibitem[Feldman \& Laming(2000)]{Feldman2000} Feldman, U., Laming, J. M. 2000, \solphys, 61, 222
\bibitem[Fisher(1989)]{Fisher1989} Fisher, G. H. 1989, \apj, 346, 1019
\bibitem[Fisher et al.(1985a)]{Fisher1985a} Fisher, G. H., Canfield, R. C., \& McClymont, A. N. 1985a, \apj, 289, 414
\bibitem[Fisher et al.(1985b)]{Fisher1985b} Fisher, G. H., Canfield, R. C., \& McClymont, A. N. 1985b, \apj, 289, 425
\bibitem[Fisher et al.(1985c)]{Fisher1985c} Fisher, G. H., Canfield, R. C., \& McClymont, A. N. 1985c, \apj, 289, 434
\bibitem[Fletcher et al.(2011)] {Fletcher2011} Fletcher, L., Dennis, B. R., Hudson, H. S., Krucker, S., Phillips, K., Veronig, A., Battaglia, M., Bone, L., Caspi, A., Chen, Q., Gallagher, P., Grigis, P. T., Ji, H., Liu, W., Milligan, R. O., \& Temmer, M. 2011, SSRv, 159, 19
\bibitem[Grigis \& Benz(2004)]{Grigis2004} Grigis, P. C., Benz, A. O. 2004, A\&A, 426, 1093
\bibitem[Hannah et al.(2008)]{Hannah2008} Hannah, I. G., Christe, S., Krucker, S., Hurford, G. J., Hudson, H. S., \& Lin, R. P., 2008, \apj, 677, 704
\bibitem[Heinzel et al.(1994)]{Heinzel1994} Heinzel, P., Karlicky, M., Kotrc, P., \& Svestka, Z., 1994, \solphys, 152, 393
\bibitem[Holman et al.(2003)]{Holman2003} Holman, G. D., Sui, L., Schwartz, R. A., \&  Emslie, A. G. 2003, \apj, 595, L97
\bibitem[Holman et al.(2011)]{Holman2011} Holman, G. D., Aschwanden, M. J., Aurass, H., Battaglia, M., Grigis, P. C., Kontar, E. P., Liu, W., Saint-Hilaire, P., \& Zharkova, V. V. 2011, SSRv, 159, 107
\bibitem[Hurford et al.(2002)]{Hurford2002} Hurford, G. J., Schmahl, E. J., Schwartz, R. A., et al. 2002, \solphys, 210, 61
\bibitem[Hurford(2004)]{Hurford2004} Hurford, G. J., 2004, Private communication
\bibitem[Kasparova \& Heinzel(2002)]{Kasparova2002} Kasparova, J. \& Heinzel, P. 2002, A\&A, 382, 688
\bibitem[Kasparova et al.(2009)]{Kasparova2009} Kasparova, J., Varady, M., Heinzel, P., Karlicky, M., \& Moravec, Z. 2009, A\&A, 499, 923
\bibitem[Klimchuk et al.(2008)]{Klimchuk2008} Klimchuk, J.~A., Patsourakos, S., \& Cargill, P.~J. 2008, \apj, 682, 1351
\bibitem[Kontar \& Jeffrey(2010)]{Kontar2010} Kontar, E. P., \& Jeffrey, N. L. S., 2010, A\&A, 513, L2, 4 pp.
\bibitem[Landi et al.(2006)]{Landi2006} Landi, E., Del Zanna, G., Young, P. R., Dere, K. P.,  Mason, H. E., \& Landini M. 2006, ApJS, 162, 261
\bibitem[Lemen et al.(2012)]{Lemen2012} Lemen, J.R., Title, A.M., Akin, D.J., Boerner, P.F., Chou, C., et al. 2012, \solphys, 275, 17
\bibitem[Lin et al.(2002)] {Lin2002} Lin, R. P., Dennis, B. R., Hurford, G. J., Smith, D. M., Zehnder, A., et al. 2002, \solphys, 210, 3
\bibitem[Liu, Petrosian, \& Mariska(2009)]{Liu2009} Liu, W., Petrosian, V., \&  Mariska, J. T. 2009, \apj, 702, 1553
\bibitem[Liu et al.(2013)]{Liu2013} Liu, W., Qiu, J., Longcope, D. W., \& Caspi, A. 2013, \apj, 770, 111
\bibitem[Lucy(1974)]{Lucy1974} Lucy, L. B. 1974, \apj, 79, 745
\bibitem[Mariska et al.(1982)]{Mariska1982} Mariska, J. T., Boris, J. P., Oran, E. S., Young, T. R. Jr., \& Doschek, G. A. 1982, \apj, 255, 738
\bibitem[Mariska \& Poland(1985)]{Mariska1985} Mariska, J. T., \& Poland, A. I. 1985, \solphys, 96, 317
\bibitem[Mariska et al.(1989)]{Mariska1989} Mariska, J. T., Emslie, A. G., \& Li, P. 1989, \apj, 341, 1067
\bibitem[Mazzotta et al.(1998)]{Mazzotta1998} Mazzotta, P., Mazzitelli, G., Colafrancesco, S., \& Vittorio, N. 1998, A\&AS, 133, 403
\bibitem[Mein(1991)]{Mein1991} Mein, P. 1991, A\&A, 248, 669
\bibitem[McTiernan \& Petrosian(1990)]{McTiernan1990} McTiernan, J. M., \& Petrosian, V. 1990, \apj, 359, 524
\bibitem[Metcalf et al.(1996)]{Metcalf1996} Metcalf, T.R., Hudson, H.S., Kosugi, T., Puetter, R. C., \& Pina, R. K. 1996, \apj, 466, 585
\bibitem[Radziszewski \& Rudawy(2013)]{Radziszewski2013} Radziszewski, K., Rudawy, P. 2013, \solphys, 284, 397
\bibitem[Radziszewski et al.(2011)]{Radziszewski2011} Radziszewski, K., Rudawy, P., Phillips, K.J.H., 2011, A\&A, 535, A123
\bibitem[Radziszewski et al.(2007)]{Radziszewski2007} Radziszewski, K., Rudawy, P., Phillips, K.J.H., 2007, A\&A, 461, 303
\bibitem[Radziszewski et al.(2006)]{Radziszewski2006} Radziszewski, K., Rudawy, P., Phillips, K.J.H., Dennis, B.R., 2006, Adv. Space Res. 37, 1317
\bibitem[Reale et al.(1997)]{Reale1997} Reale, F., Betta, R., Peres, G., Serio, S., \& McTiernan, J. 1997, A\&A, 325, 782
\bibitem[Reep et al.(2013)]{Reep2013} Reep, J. W., Bradshaw, S. J., \& McAteer, R. T. J. 2013, \apj, 778, 76
\bibitem[Reep et al.(2016)]{Reep2016} Reep, J. W., Bradshaw, S. J., \& Holman, G. D. 2016, \apj, 818, 44
\bibitem[Richardson(1972)]{Richardson1972} Richardson, W. H. 1972, JOSA 62, 55
\bibitem[Rompolt et al.(1994)]{Rompolt1994} Rompolt, B., Mein, P., Mein, N., Rudawy, P., Berlicki, A., 1994, JOSO Annual Report 1993, 87
\bibitem[Scherrer et al.(2012)]{Scherrer2012} Scherrer, P.H., Schou, J., Bush, R.I., Kosovichev, A.G., Bogart, R.S., et al. 2012, \solphys, 275, 207
\bibitem[Schmeltz et al.(1994)]{Schmeltz1994} Schmelz, J.T., Holman, G.D., Brosius, J.W., \& Willson, R. F. 1994, \apj, 434, 786
\bibitem[Serio et al.(1991)]{Serio1991} Serio, S., Reale, F., Jakimiec, J., Sylwester, B., \& Sylwester, J. 1991, A\&A, 241, 197
\bibitem[Siarkowski et al.(2009)]{Siarkowski2009} Siarkowski, M., Falewicz, R., \& Rudawy, P. 2009, ApJL, 705, L143
\bibitem[Sui et al.(2007)]{Sui2007} Sui, L. H., Holman, G. D., \& Dennis, B. R. 2007, \apj, 670, 862
\bibitem[Tandberg-Hanssen \& Emslie(1988)]{Tandberg1988} Tandberg-Hanssen, E., \& Emslie, A. G., 1998, in The physics of solar flares, Cambridge and New York, Cambridge University Press, p. 113
\bibitem[Vernazza, Avrett \& Loeser(1973)]{Vernazza1971} Vernazza, J. E., Avrett, E. H., Loeser, R. 1973, \apj, 184, 605
\bibitem[Vernazza, Avrett \& Loeser(1981)]{Vernazza1981} Vernazza, J. E., Avrett, E. H., Loeser, R. 1971, \apjs, 45, 635
\bibitem[Warmuth et al.(2009a)]{Warmuth2009a} Warmuth, A., Holman, G. D., Dennis, B. R., Mann, G., Aurass, H., \& Milligan, R. O. 2009a, \apj, 699, 917
\bibitem[Warmuth et al.(2009b)]{Warmuth2009b} Warmuth, A., Mann, G., \& Aurass, H. 2009b, A\&A, 494, 677
\bibitem[Warren(2006)]{Warren2006} Warren, H. P. 2006, \apj, 637, 522
\bibitem[White et al.(2005)]{White2005} White, S., Thomas, R. J., \& Schwartz, R. A. 2005, \solphys, 227, 231

\end{thebibliography}
\end{document}